\documentclass[aps,prd,onecolumn,10pt]{revtex4-1}

\usepackage{graphicx}
\usepackage{amsmath}
\usepackage{hyperref}

\newcommand{\be}{\begin{equation}}
\newcommand{\ee}{\end{equation}}
\newcommand{\bea}{\begin{eqnarray}}
\newcommand{\eea}{\end{eqnarray}}
\newcommand{\nn}{\nonumber}

\begin{document}
\title{Next-to-leading order gravitational spin-orbit coupling \\ in an effective field theory approach}
\author{Michele Levi}
\email{michele@phys.huji.ac.il}
\affiliation{Racah Institute of Physics, Hebrew University, Jerusalem 91904, Israel}
\date{\today}

\begin{abstract} 
We use an effective field theory (EFT) approach to calculate the next-to-leading order (NLO) gravitational spin-orbit interaction between two spinning compact objects. The NLO spin-orbit interaction provides the most computationally complex sector of the NLO spin effects, previously derived within the EFT approach. In particular, it requires the inclusion of non-stationary cubic self-gravitational interaction, as well as the implementation of a spin supplementary condition (SSC) at higher orders. The EFT calculation is carried out in terms of the nonrelativistic gravitational field parametrization, making the calculation more efficient with no need to rely on automated computations, and illustrating the coupling hierarchy of the different gravitational field components to the spin and mass sources. Finally, we show explicitly how to relate the EFT derived spin results to the canonical results obtained with the Arnowitt-Deser-Misner (ADM) Hamiltonian formalism. This is done using noncanonical transformations, required due to the implementation of covariant SSC, as well as canonical transformations at the level of the Hamiltonian, with no need to resort to the equations of motion or the Dirac brackets.
\end{abstract}

\maketitle	

\section{Introduction}

For some years now major efforts have been undertaken in order to detect gravitational radiation, one of Einstein's most important predictions of the theory of general relativity (GR), longstanding since 1916. Currently, three ground-based gravitational wave (GW) detectors are operating worldwide: LIGO (Laser Interferometer Gravitational-wave Observatory), Virgo, and GEO600 \cite{ligo, virgo, geo600}. More advanced ground-based interferometers such as AIGO (Australian International Gravitational Observatory), that will complement the existing network of detectors due to its location in Australia \cite{aigo}, or LCGT (Large-scale Cryogenic Gravitational-wave Telescope) in Japan \cite{lcgt}, are expected to start operating in the near future. Moreover, LISA (Laser Interferometer Space Antenna) is expected to be the first space-based GW detector, exploring a complementary range of signal frequencies \cite{lisa}.

One of the most promising sources of GW are binary systems of compact objects, in particular of black holes or neutron stars, that are found in the inspiral phase of their evolution. At this stage the dynamics of the binary can be described analytically in terms of the post-Newtonian (PN) approximation of GR \cite{Blanchet:2002av}. In order to succeed in the detection of GW signals, which is performed by matched filtering, accurate theoretical waveform templates of the inspiral are required, and so PN corrections should be obtained to high order, at least up to 3PN. The reason is that it is the overall value of the contribution of the PN correction to the phase of the waveform that matters for detection, and not its relative value, which can be numerically small with respect to the leading contribution. Since astrophysical objects are expected to have a \textit{spin}, gravitational spin effects must be considered in the description of binary dynamics. Indeed, the leading order (LO) gravitational spin effect in compact binaries is evaluated to contribute already at 1.5PN \cite{Tulczyjew:1959a,Barker:1975ae}. Spin effects may enhance GW signals and increase the event rates for their detection \cite{Arun:2008kb,Tessmer:2010hp}. 

A substantial possibility for progress in the analytical treatment was given when a novel effective field theory (EFT) approach for treating the PN formalism of GR was introduced by Goldberger and Rothstein \cite{Goldberger:2004jt}. The EFT approach is very advantageous in applying the efficient standard tools of quantum field theory to GR, notably handling the regularization required for higher order corrections in the PN approximation with the standard renormalization scheme. Moreover, this approach is appropriate to handle various problems that require the treatment of several typical length scales. Originally, the EFT approach was suggested for implementation on the inspiral phase of a binary to yield predictions of gravitational radiation \cite{Goldberger:2004jt,Goldberger:2007hy}. Then, it was used to obtain the thermodynamics of higher dimensional Kaluza-Klein (KK) black holes \cite{Chu:2006ce} and was extended for the rotating \cite{Kol:2007rx} and charged \cite{Gilmore:2009ea} cases. 

Various applications of the EFT approach then followed. Next-to-leading order (NLO) PN spin1-spin2 and spin-squared effects in inspiralling compact binaries were calculated for the first time \cite{Porto:2008tb,Levi:2008nh,Porto:2008jj} (the complete results were also obtained with the Arnowitt-Deser-Misner (ADM) Hamiltonian approach in \cite{Steinhoff:2007mb,Steinhoff:2008ji,Hergt:2008jn,Hergt:2010pa}). The 2PN dynamics for binaries were reproduced \cite{Gilmore:2008gq}, and new results for n-body systems were obtained with automated calculations \cite{Chu:2008xm}. The possibility to extract the three- and four-graviton vertices from binary pulsars and coalescing binaries was examined in \cite{Cannella:2009he}. Radiative corrections in compact binaries were reproduced up to 3PN order \cite{Goldberger:2009qd}. Self-force on extreme mass ratio inspirals and radiation reaction were also treated in the EFT approach \cite{Galley:2008ih,Galley:2009px}. Most recently even finite size corrections to the radiation reaction in classical \textit{electrodynamics} were found using the EFT approach \cite{Galley:2010es}.

In this work, we make an application of the EFT approach to the calculation of the NLO spin-orbit (SO) interaction in inspiralling compact binaries. This is the only NLO PN spin correction that was not handled before in the EFT approach due to its higher complexity. In particular, the NLO SO sector requires the inclusion of nonstationary cubic self-gravitational interaction, as well as the implementation of a spin supplementary condition (SSC) at higher orders. Here, we work with the reduction over the time dimension that was proposed within the EFT approach by \cite{Kol:2007bc}, and use the nonrelativistic parametrization of the gravitational field, which was first utilized in the computation of the NLO spin1-spin2 interaction \cite{Levi:2008nh}. The nonrelativistic gravitational (NRG) field variables prove to greatly simplify the calculation and make it feasible with no need for automated computations. The NRG fields also illustrate the coupling hierarchy of the different gravitational field components to the spin and mass sources. We make an effort and put an emphasis on presenting the treatment of self-gravitating spinning objects in an EFT approach in a clear and self-contained manner. Then, after we give a detailed account of the EFT calculation of the NLO SO sector, we use noncanonical transformations, required due to the implementation of covariant SSC, as well as canonical transformations at the level of the Hamiltonian. This leads us to the canonical result calculated in the ADM Hamiltonian formalism \cite{Damour:2007nc}, which also agrees with that of \cite{Faye:2006gx}, whose authors first obtained the complete NLO SO interaction equations of motion (EOM). Naturally, it is more efficient to use variable transformations and remain in the level of the Lagrangian/Routhian, or Hamiltonian, than to resort to the treatment of EOM and the Dirac brackets. 

Throughout this paper, we use $c\equiv1$, $\eta_{\mu\nu}\equiv diag[1,-1,-1,-1]$, and the convention for the Riemann tensor is 
$R^\mu_{~\nu\alpha\beta}\equiv\partial_\alpha\Gamma^\mu_{\nu\beta}-\partial_\beta\Gamma^\mu_{\nu\alpha}+\Gamma^\mu_{\lambda\alpha}\Gamma^\lambda_{\nu\beta}-\Gamma^\mu_{\lambda\beta}\Gamma^\lambda_{\nu\alpha}$. 
Greek letters will denote indices in the global coordinate frame. Lowercase Latin letters from the beginning of the alphabet will denote indices in the local Lorentz frame, while uppercase Latin letters denote the body fixed, corotating Lorentz frame. All these indices will run over 0 to 3, while spatial tensor indices over 1 to 3, will be denoted with lowercase Latin letters from the middle of the alphabet. The letter t may be used as an alternate for the time index 0. We will also use the following notation $\int_{\bf{k}} \equiv \int \frac{d^3{\bf{k}}}{(2\pi)^3}$ for abbreviation (boldface characters are used to denote 3-vectors). 

The paper is organized as follows. In Sec.~\ref{sec:eftKK} we review the EFT approach for the binary inspiral problem in the PN approximation, with the NRG field parametrization, and present the Feynman rules in the linearized harmonic gauge, required for the EFT computation. In Sec.~\ref{sec:spineft} we present in a self-contained manner the formalism for the treatment of self-gravitating spinning objects in an EFT approach and present the Feynman rules involving the spin required for the computation with the NRG fields. In Sec.~\ref{sec:solo} we derive the LO SO potential, which is required to obtain the NLO SO Hamiltonian, but also to illustrate the most simple EFT calculation in the SO sector and its ambiguities. In Sec.~\ref{sec:csonlo} we present in detail the full calculation of the NLO SO Lagrangian/Routhian, going over all contributing Feynman diagrams, from one- and two-graviton exchanges to cubic self-gravitational interaction diagrams. In Sec.~\ref{sec:sonlo} we present the NLO SO Lagrangian/Routhian EFT result and derive the NLO SO Hamiltonian, using noncanonical as well as canonical transformations to obtain the canonical result in \cite{Damour:2007nc} in the ADM gauge. In Sec.~\ref{sec:concl} we summarize our main conclusions. 

We note that after this work was complete, there appeared \cite{Perrodin:2010dy,Porto:2010tr}, which study the same problem.

\section{EFT approach for binary inspiral in the PN approximation}\label{sec:eftKK}

As explained in the EFT description of the binary inspiral problem \cite{Goldberger:2004jt}, expanding the metric around flat spacetime, the metric can be decomposed into potential and radiation modes, both with the same typical time variation scale of the binary orbital frequency $v/r$, but with a different typical length scale for each, $r$ and $r/v$, respectively, where $v$ and $r$ are the typical orbital parameters of the binary, and we are working in the limit $v\ll1$ of the PN approximation. Thus, as far as the potential gravitons are concerned, the radiation component of the field is just a slowly varying background field of soft momenta gravitons. Moreover, considering these typical scales, we see that the potential gravitons are off shell with their frequency being much smaller than their momentum, and hence can be approximated as stationary to leading order. This motivates a KK reduction over the time dimension for the potential field modes as suggested in \cite{Kol:2007rx,Kol:2007bc}. Note that since we are concerned with conservative dynamics in this work, we only need the potential modes anyway and can simply set the radiation modes to zero.

Therefore, instead of the common Lorentz covariant metric parametrization in the weak field limit, of the form $g_{\mu\nu} = \eta_{\mu\nu} + h_{\mu\nu}$, the metric is parametrized in a nonrelativistic form according to the Kaluza-Klein ansatz
\be \label{eq:kka}
d\tau^2 = g_{\mu\nu}dx^{\mu}dx^{\nu} \equiv e^{2 \phi}(dt - A_i\, dx^i)^2 -e^{-2 \phi} \gamma_{ij}dx^i dx^j~.
\ee
This defines a set of new fields $(\phi, A_i, \gamma_{ij})$, with a scalar field $\phi$ corresponding to the Newtonian potential, a 3-vector field $A_i$, corresponding to the so-called gravito-magnetic vector, and a 3-dimensional symmetric 2-tensor field $\gamma_{ij}$, the nonrelativistic gravitational fields as discussed in \cite{Kol:2007bc}. Then, in terms of the NRG fields the metric reads
\be \label{eq:gkk}
g_{\mu\nu}=
\left(\begin{array}{cc} 
e^{2\phi}      & -e^{2\phi} A_j \\
-e^{2\phi} A_i & -e^{-2\phi}\gamma_{ij}+e^{2\phi} A_i A_j
\end{array}\right)\simeq
\left(\begin{array}{cc} 
1+2\phi+2\phi^2 & -A_j-2A_j\phi \\
-A_i-2A_i\phi   & -\delta_{ij}+2\phi\delta_{ij}-\sigma_{ij}-2\phi^2\delta_{ij}+2\phi\sigma_{ij}+A_iA_j
\end{array}\right),
\ee
where $\gamma_{ij}\equiv\delta_{ij}+\sigma_{ij}$, and we have written the approximation for the metric in the weak field limit up to second order in the fields, which is the order required in this work. In the ground state $\langle\phi\rangle=\langle A_i\rangle=\langle\sigma_{ij}\rangle=0$, and the metric reduces to Minkowski spacetime. We should also have the form of the inverse metric, and it is given by
\be
g^{\mu\nu}=
\left(\begin{array}{cc} 
e^{-2\phi}-e^{2\phi}\gamma^{ij}A_iA_j & -e^{2\phi} A^j \\
-e^{2\phi} A^i                        & -e^{2\phi}\gamma^{ij}
\end{array}\right)\simeq
\left(\begin{array}{cc} 
1-2\phi+2\phi^2-A_kA_k         & -A_j-2\phi A_j+\sigma_{jk}A_k \\
-A_i-2\phi A_i+\sigma_{ik}A_k  & -\delta_{ij}-2\phi\delta_{ij}+\sigma_{ij}-2\phi^2\delta_{ij}+2\phi\sigma_{ij}-\sigma_{il}\sigma_{lj}
\end{array}\right),
\ee
where $\gamma^{ij}\gamma_{jk}\equiv\delta_{ik}$, i.e.~$\gamma^{ij}$ is the inverse of $\gamma_{ij}$, and $A^i\equiv\gamma^{ij}A_j$. Here too, we have written the weak field limit of the inverse metric up to second order in the fields.

Next, we consider the purely gravitational action. It is the usual Einstein-Hilbert (EH) action plus a gauge fixing term of our choice
\be
S_g = S_{EH} + S_{GF} = -\frac{1}{16\pi G} \int d^4x \left(\sqrt{g} R+ \cal{L}_{\it{GF}}\right).
\ee
We suppress the time dependence of the fields in the EH action to obtain the KK reduced action for the gravitational field, given by
\be
S_{KK} \equiv S_{EH}|_{stat.} = -\frac{1}{16\pi G} \int dt d^3x ~\sqrt{\gamma} \left[ -R[\gamma_{ij}] + 2\gamma^{ij} \partial_i \phi\partial_j \phi  -  \frac{1}{4} e^{4\phi} F_{ij} F_{kl}\gamma^{ik}\gamma^{jl}\right]~, 
\ee  
where $\gamma \equiv det(\gamma_{ij})$ and $F_{ij}\equiv\partial_i A_j-\partial_j A_i$. 

Now, we must include a gauge fixing term to set the gravitational action completely. We should be able to invert the graviton kinetic term, i.e.~the quadratic part of the Lagrangian, in order to obtain the graviton propagators. Then, the propagators can be read off from the terms that are quadratic in the fields, depending on the choice of gauge fixing term. Quadratic terms with time derivatives are suppressed as subleading corrections in powers of $v^2$: We recall that the potential gravitons are instantaneous within the leading stationary approximation, representing off shell gravitons. A time derivative adds a factor of $v$ for potential modes, hence the propagator corrections are suppressed by $v^2$. The quadratic part of the KK reduced action (after two integrations by parts) is given by 
\be 
S_{KK[g^2]} = -\frac{1}{32\pi G} \int dt d^3x \left[\frac{1}{2}(\partial_k\sigma_{ij})^2 -(\partial_j\sigma_{ij})^2+\partial_j\sigma_{ij}\partial_i\sigma_{kk}-\frac{1}{2}(\partial_i\sigma_{jj})^2 + 4 (\partial_i \phi)^2 - (\partial_iA_j)^2+(\partial_iA_i)^2\right]. 
\ee
Hence, we take the natural gauge adequate for the NRG fields, namely, the Lorentz gauge for the vector field $A_i$, and the harmonic gauge for the 3-dimensional 2-tensor field $\sigma_{ij}$. This is just equivalent to the harmonic gauge for the Lorentz covariant parametrization of the metric. Thus, the gauge fixing term is given by
\be\label{eq:gf}
S_{\it{GF}[KK]} = \frac{1}{32\pi G} \int dtd^3x\left[\left(\partial_{i}A_i\right)^2 - \left(\partial_j\sigma_{ij}-\frac{1}{2}\partial_i\sigma_{jj}\right)^2\right]. 
\ee
Therefore, the NRG scalar, vector, and 2-tensor field propagators in the harmonic gauge are given by 
\begin{align}
\label{eq:prphi} \parbox{18mm}{\includegraphics{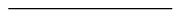}}
 & = \langle{~\phi(x_1)}~~{\phi(x_2)~}\rangle =~~~ \frac{1}{8}~~~~ \int_{\bf{k}} \frac{e^{i{\bf k}\cdot\left({\bf x}_1 - {\bf x}_2\right)}}{{\bf k}^2}~\delta(t_1-t_2),\\ 
\label{eq:prA} \parbox{18mm}{\includegraphics{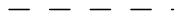}}
 & = \langle{A_i(x_1)}~{A_j(x_2)}\rangle = -\frac{1}{2}~\delta_{ij} \int_{\bf{k}} \frac{e^{i{\bf k}\cdot\left({\bf x}_1 - {\bf x}_2\right)}}{{\bf k}^2}~\delta(t_1-t_2),\\ 
\label{eq:prsigma} \parbox{18mm}{\includegraphics{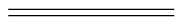}}
 & = \langle{\sigma_{ij}(x_1)}{\sigma_{kl}(x_2)}\rangle =~ P_{ij;kl}~ \int_{\bf{k}}\frac{e^{i{\bf k}\cdot\left({\bf x}_1 - {\bf x}_2\right)}}{{\bf k}^2}~\delta(t_1-t_2),
\end{align}
where $P_{ij;kl}\equiv\frac{1}{2}\left(\delta_{ik}\delta_{jl}+\delta_{il}\delta_{jk}-2\delta_{ij}\delta_{kl}\right)$. Here and henceforth, the Feynman rules are presented in position space. Note the simple form of the propagators obtained with the NRG field parametrization and harmonic gauge, in particular, that of the $\phi$ and $A_i$ fields, which dominate in the interaction, an advantageous feature of the NRG field variables.

However, as already noted there are corrections to the nonrelativistic instantaneous graviton propagators, i.e.~to the graviton kinetic terms in the KK reduced action. These are included systematically as perturbative corrections, where each propagator correction is suppressed by a power of $v$ for each time derivative it contains. It is a departure from the stationarity approximation in the gravitational action, thus where the use of the KK action becomes insufficient, and we are bound to go back and consider the full EH action. Extracting from the EH action the quadratic time dependent terms, which contribute to the order considered here, namely, the terms which contain the scalar field $\phi$ as well as the vector field $A_i$, we have 
\be\label{eq:2eht}
S_{EH[g^2]} \supset -\frac{1}{16\pi G}\int d^4x \left[ 6(\partial_0\phi)^2 
+4\partial_iA_i\partial_0\phi +\partial_0A_i\partial_j \sigma_{ij}-\frac{1}{2}\partial_0A_i\partial_i \sigma_{jj}-\frac{1}{2}\partial_iA_i\partial_0 \sigma_{jj}-2\partial_0\phi\partial_0\sigma_{ii} \right].
\ee
Thus in order to simplify as much as possible, we refine the gauge fixing term of Eq.~(\ref{eq:gf}) to eliminate undesired quadratic vertices. Hence, we reset it to
\be
S_{\it{GF}} = \frac{1}{32\pi G} \int d^4x\left[\left( \partial_{i}A_i+\left(4\partial_0\phi -\frac{1}{2}\partial_0\sigma_{ii}\right)\right)^2 - \left(\left(\partial_j\sigma_{ij}-\frac{1}{2}\partial_i\sigma_{jj}\right)-\partial_0A_i\right)^2\right], 
\ee
so that to the order considered here, we are left with the following time dependent quadratic terms for the scalar and vector fields 
\be\label{eq:2gt} 
(S_{EH[g^2]}+S_{GF})\supset -\frac{1}{32\pi G}\int d^4x~\bigl[(\partial_0A_i)^2 - 4(\partial_0\phi)^2\bigr].
\ee
Again, this is just equivalent to the harmonic gauge for the Lorentz covariant parametrization of the metric. Note that here we face the major advantage of the NRG parametrization in harmonic gauge -- the 2-point functions between the three different fields are 0: $\langle \phi A_i \rangle = \langle \phi~\sigma_{jk} \rangle = \langle A_i\sigma_{jk} \rangle=0$. Thus, the Feynman rules for the propagator correction vertices are given by
\begin{align}
\label{eq:prtphi}  \parbox{18mm}{\includegraphics{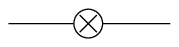}}
 & = ~~\frac{1}{8\pi G}~~\int d^4x~[\partial_0\phi(x)]^2, \\ 
\label{eq:prtA}   \parbox{18mm}{\includegraphics{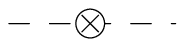}}
 & = -\frac{1}{32\pi G} \int d^4x~[\partial_0A_i(x)]^2. 
\end{align}
The encircled crosses represent the self-gravitational vertices that contain two time derivatives.

We will also have contributions from three-graviton vertices of cubic gravitational self-interaction. These are extracted from the purely gravitational action as we are using the linearized harmonic gauge, and not the fully harmonic gauge, i.e.~since our gauge fixing is quadratic in the fields, we extract the cubic part from the KK or EH action alone. The LO three-graviton vertices scale as $v^2$ according to the EFT power counting \cite{Goldberger:2004jt}, and they are easily derived from the cubic part of the KK reduced action. To the order considered here, there are two such cubic vertices contributing, the $\phi F^2$ and the $\sigma \phi^2$ vertices given by
\be\label{eq:3kk}
S_{KK[g^3]} \supset -\frac{1}{16\pi G}\int dt d^3x~\bigl[\left(-2\phi \left(\partial_iA_j\partial_iA_j-\partial_iA_j\partial_jA_i\right)\right) + \left(-2\sigma_{ij}\partial_i\phi\partial_j\phi+\sigma_{jj}\partial_i\phi\partial_i\phi\right)\bigr].
\ee
However, at the order considered, we also have a contribution from another three-graviton vertex, including yet again a departure from the KK stationary action. This vertex contains a single time derivative, hence it scales as $v^3$. Therefore, to obtain this vertex we have to go through the tedious extraction of the cubic part of the EH action, containing a very large number of terms, each with a complicated tensor index structure as was done in \cite{Goldberger:2004jt}. The contributing time dependent cubic vertex is the $A \phi^2$ vertex given by
\be\label{eq:3eh}
S^{g^3}_{EH} \supset -\frac{1}{16\pi G}\int dt d^3x~[4A_i\partial_i\phi\partial_0\phi]~.
\ee                                                                    
Hence, the Feynman rules for the three-graviton vertices are given by
\begin{align}
\label{eq:phiA^2} \parbox{18mm}{\includegraphics{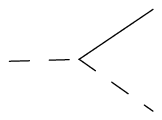}}
 & = ~~\frac{1}{8\pi G}\int d^4x~\biggl[\phi(x)\partial_iA_j(x)\biggl(\partial_iA_j(x)-\partial_jA_i(x)\biggr)\biggr], \\ 
\label{eq:sigmaphi^2} \parbox{18mm}{\includegraphics{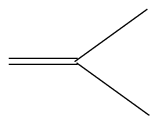}}
 & = ~ \frac{1}{16\pi G}\int  d^4x~[2\sigma_{ij}(x)\partial_i\phi(x)\partial_j\phi(x)-\sigma_{jj}(x)\partial_i\phi(x)\partial_i\phi(x)], \\ 
\label{eq:Aphi^2t} \parbox{18mm}{\includegraphics{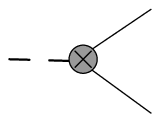}}
 & = -\frac{1}{4\pi G}\int d^4x~[A_i(x)\partial_i\phi(x)\partial_0\phi(x)].
\end{align}
The gray encircled cross represents the self-gravitational vertex, containing a single time derivative.
 
Now we consider the gravitational coupling to the two massive compact objects. We take the worldline action of a point particle (PP) for each of the two objects, so that to the order considered here we have for the binary system 
\be
S_{pp(m)} = -\sum_{n=1}^2 \int m_n d\sigma_n + \cdots,
\ee
where the ellipsis denotes subleading operators encoding finite size effects. Here, we do not consider them since they enter only at 5PN \cite{Goldberger:2004jt}. We parametrize the worldline using the coordinate time $t=x^0$, i.e.~$\sigma=t$, so that we have for $u^\mu\equiv dx^\mu/d\sigma$: $u^0=1$, $u^i=dx^i/dt\equiv v^i$. Expanding the PP action of a massive object using Eq.~(\ref{eq:kka}), up to the order required here in the fields and velocities, gives rise to the following couplings to the worldline mass: 
\bea \label{eq:lppm}
S_{pp(m)} &&= -\int m d\sigma = -m \int dt \sqrt{g_{\mu\nu}\frac{dx_\mu}{dt} \frac{dx_\nu}{dt}} = -m \int dt~    \left[e^{\phi}\sqrt{\left(1-A_iv^i\right)^2-e^{-4\phi}\gamma_{ij}v^iv^j}~\right] \nn \\ 
    &&= -m \int dt~\left[1-\frac{1}{2}v^2+\phi-A_iv^i+\frac{3}{2} \phi v^2-\frac{1}{2}\sigma_{ij} v^iv^j-\frac{1}{2}A_iv^i v^2+\cdots+\frac{1}{2}\phi^2-\phi A_iv^i+\cdots\right],
\eea
where the ellipsis stands for one-, two-, and n-graviton couplings, and velocity powers beyond the order considered here. It is clear that any number of gravitons can couple to the worldline mass. Moreover, the LO couplings of $\phi$, $A_i$, and $\sigma_{ij}$ are $O(v^0)$, $O(v^1)$, and $O(v^2)$, respectively, and all couplings contain an infinite power series in $v^2$. The Feynman rules for the one-graviton couplings to the worldline mass are given by
\begin{align}
\label{eq:mphi} \parbox{12mm}{\includegraphics{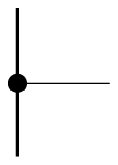}}
 & = - m \int dt~\phi({\bf{x}}(t))~\left[1+\frac{3}{2}v(t)^2+\cdots\right], \\ 
\label{eq:mA} \parbox{12mm}{\includegraphics{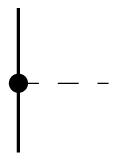}}
 & = ~m \int dt~A_i({\bf{x}}(t))v^i(t)~\left[1+\frac{1}{2}v(t)^2+\cdots\right], \\ 
\label{eq:msigma} \parbox{12mm}{\includegraphics{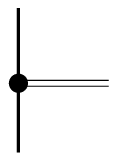}}
 & = ~m \int dt~\sigma_{ij}({\bf{x}}(t))v^i(t)v^j(t)~\left[\frac{1}{2}+\cdots\right],
\end{align}
where the heavy solid lines represent the worldlines, and the spherical black blobs represent the masses on the worldline. Again, the ellipsis denotes higher orders in $v$, beyond the order considered here. For the two-graviton couplings to the worldline mass, we have the following Feynman rules:   
\begin{align}
\label{eq:mphi^2}  \parbox{12mm}{\includegraphics{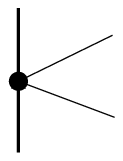}}
 & = - m \int dt~\phi({\bf{x}}(t))^2~\left[\frac{1}{2}+\cdots\right], \\ 
\label{eq:mphiA}   \parbox{12mm}{\includegraphics{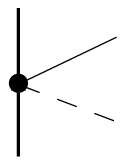}}
 & = ~m \int dt~\phi({\bf{x}}(t))A_i({\bf{x}}(t))v^i(t)~\biggl[1+\cdots\biggr].
\end{align}

\section{Self-gravitating spinning objects with an EFT approach} \label{sec:spineft}

The interaction of classical spin with gravity was discovered in 1937 by Mathisson \cite{Mathisson:1937zz}. However, the EOM for a spinning test body in GR were derived in their modern form by Papapetrou \cite{Papapetrou:1951pa}. The Mathisson-Papapetrou (MP) EOM were obtained based on a multipole formalism for extended bodies and describe the motion of a spinning body in the so-called pole-dipole approximation, where quadrupole and higher multipole moments are neglected. Later, the MP equations were rederived using an action approach. For example, Bailey and Israel obtained the MP equations from a variation of the action, though they did not specify the Lagrangian form \cite{Bailey:1975fe}, similar to the analogous treatment given in flat spacetime by Hanson and Regge \cite{Hanson:1974qy}. Yee and Bander used a Routhian, whose form they specified, to obtain the EOM \cite{Yee:1993ya}. More recently, further similar treatments using an action approach were made in \cite{Porto:2008tb,Porto:2005ac,Steinhoff:2009ei,Barausse:2009aa}. Here, we consider the self-gravitating spinning objects of the binary system as point particles in what is equivalent to the pole-dipole approximation, i.e.~we only work to linear order in the particles spins. That is ,we are not concerned with finite size effects, which are quadratic in the spin, such as self-induced effects or tidal deformations, or in any intrinsic permanent multipole moment beyond spin. 

We start by defining a tetrad field as a set consisting of a timelike future-oriented vector and three spacelike vectors, which satisfy
\be \label{eq:ttrd}
g_{\mu\nu}e^\mu_ae^\nu_b = \eta_{ab},
\ee
with $\eta_{ab} \equiv diag[1,-1,-1,-1]$ the flat space Minkowski metric, and $e^\mu_a$ the set of vectors with the tetrad index $a$ ranging from 0 to 3. Tetrad indices are raised and lowered with the Minkowski metric $\eta_{ab}$, e.g.~$e^{a\mu}\equiv\eta^{ab}e^\mu_b$. This is equivalent to defining the reciprocal tetrad field by
\be \label{eq:rttrd}
g_{\mu\nu}e^{a\mu} e_b^\nu = \delta^a_b, 
\ee
which also tells us that the tetrad is an orthonormal set. Now, one can easily obtain the completeness relation 
\be \label{eq:cttrd}
\eta_{ab}e^{a\mu} e^b_\nu = \delta^\mu_\nu.
\ee
From Eqs.~(\ref{eq:ttrd})-(\ref{eq:cttrd}) it is easy to obtain the following expressions for the metric and its inverse in terms of the tetrad fields:
\be\label{eq:gfixe}
g_{\mu\nu} = \eta_{ab}e^a_\mu e^b_\nu, \,\,\,\, g^{\mu\nu} = \eta^{ab}e^\mu_a e^\nu_b.
\ee
The projections of a vector $V^\mu$ (and similarly for a tensor of any rank) onto the internal tetrad space are defined as $V^a \equiv e^a_\mu V^\mu$, $V_a \equiv e^\mu_a V_\mu$, or conversely we have $V^\mu \equiv e^\mu_a V^a$, $V_\mu \equiv e^a_\mu V_a$.

In order to describe the rotation of a spinning object in curved spacetime, we need two different tetrads. First, we need the tetrad of the body fixed, corotating Lorentz frame, carried by the rotating particle. To begin with we would like to follow the evolution of this tetrad along the worldline. We would also require a reference tetrad that covers the background spacetime, the tetrad of the local Lorentz frame of the particle. The internal tetrad space is Lorentz invariant, so while satisfying Eq.~(\ref{eq:gfixe}), we have a freedom in the choice of this tetrad that will be fixed. We recall that our convention for the notation of the body fixed tetrad is by Latin uppercase letters from the beginning of the alphabet, whereas the background tetrad is denoted by Latin lowercase letters from the beginning of the alphabet. The body fixed tetrad is related to the background tetrad by a Lorentz transformation, i.e.~$e_A^{\mu} = \Lambda_A^a e_a^\mu$, where the Lorentz transformations and their inverse are defined by $\Lambda^{Aa}\Lambda^{Bb}\eta_{AB}=\eta^{ab}$ and $\Lambda_{Aa}\Lambda_{Bb}\eta^{ab}=\eta_{AB}$. The Lorentz transformations relating the two tetrads should reduce to pure rotations in the local rest frame. Now we can proceed to define the generalized angular velocity in curved spacetime, which describes how the body fixed tetrad rotates along the worldline, by
\be \label{eq:defomg}
\Omega^{\mu\nu} \equiv e_A^\mu\frac{De^{A\nu}}{D\sigma},
\ee
where $D/D\sigma$ is the covariant derivative with respect to the worldline parameter $\sigma$. It is easy to verify that $\Omega^{\mu\nu}$ is an antisymmetric tensor. This definition is a generalization of the definition for flat spacetime given by $\Omega^{ab}\equiv\Lambda_A^a\frac{d\Lambda^{Ab}}{d\sigma}$ \cite{Hanson:1974qy}. 

We should construct an action for the spinning particle that respects general coordinate invariance, reparametrization invariance (of the worldline parameter of the objects), as well as SO(3) residual Lorentz  invariance of the tetrad internal space. The minimal coupling part of the action would then be given by the PP action $S_{pp} = \int d\sigma\, L_{pp}(s_1,s_2,s_3,s_4)=\int d\sigma \,L_{pp}(u^\rho,\Omega^{\mu\nu})$, where the Lagrangian $L_{pp}$ is a function of the four covariant scalars $s_1,\ldots ,s_4$ that can be formed from the coordinate velocity $u^\mu \equiv dx^\mu/d\sigma$ and the generalized angular velocity $\Omega^{\mu\nu}$ given by \cite{Hanson:1974qy} 
\bea \label{eq:mcs}
s_1 &=& u^\mu u_\mu, \nn \\ 
s_2 &=& \Omega^{\mu\nu}\Omega_{\mu\nu}, \nn \\ 
s_3 &=& u^\mu \Omega_{\mu\nu} \Omega^{\nu\rho} u_\rho, \nn \\ 
s_4 &=& \Omega^{\mu\nu}\Omega_{\nu\rho}\Omega^{\rho\kappa}\Omega_{\kappa\mu}, \,\, or \,\,det(\Omega^{\mu\nu}). 
\eea
Since the action is reparametrization invariant, we require the Lagrangian $L_{pp}$ to be a homogeneous function of degree 1 in the velocities $u^\mu$ and $\Omega^{\mu\nu}$ \cite{Hanson:1974qy}. So that if we define the \textit{linear} momentum $p_\mu$, and the antisymmetric tensor $S_{\mu\nu}$ of the particle as 
\bea \label{eq:defps}
p_\mu &\equiv& -\left.\frac{\partial L}{\partial u^\mu}\right|_{\Omega},\label{eq:pd}\\
S_{\mu\nu} &\equiv& -2\left.\frac{\partial L}{\partial \Omega^{\mu\nu}}\right|_u\label{eq:sd},
\eea
we have from Euler's theorem that $L_{pp}$ is given by 
\be \label{eq:lpp}
L_{pp} = -p_\mu u^\mu - \frac{1}{2} S_{\mu\nu} \Omega^{\mu\nu}.
\ee
Note that in curved spacetime the linear momentum $p_\mu$ is different than the momentum conjugate to the coordinate velocity $u^\mu$, since $\Omega^{\mu\nu}$ depends on $u^\mu$ (e.g.~$u^\mu$ is explicit in the covariant differentiation in the definition of $\Omega^{\mu\nu}$ in Eq.~(\ref{eq:defomg})). Therefore, the differentiation in the definitions of Eqs.~(\ref{eq:pd}) and (\ref{eq:sd}) is special in that we treat $\Omega^{\mu\nu}$ as independent of $u^\mu$. The minus signs in these definitions are chosen to give the correct nonrelativistic limit. Indeed, the PP Lagrangian form of Eq.~(\ref{eq:lpp}) is just obtained from a minimal coupling of the flat spacetime action \cite{Hanson:1974qy}. 

A variation of the action with respect to the position $x^\mu$, and the tetrad field $e^\mu_A$, with fixed endpoints yields the EOM for the position and spin respectively \cite{Bailey:1975fe,Porto:2005ac}. The EOM are given by the well-known MP equations 
\bea 
\frac{Dp_\mu}{D\sigma}&=&-\frac{1}{2}R_{\mu\nu\rho\sigma}u^\nu S^{\rho\sigma},\label{eq:mp1}\\
\frac{DS_{\mu\nu}}{D\sigma}&=&\,\,\,\,\,\,\,\,p_\mu u_\nu-p_\nu u_\mu.\label{eq:mp2}
\eea
However, the description of the motion of a spinning object by the MP equations is not closed. The solution for the EOM is not unique due to redundant unphysical degrees of freedom associated with the spin that should be eliminated. Hence, we need to choose a so-called spin supplementary condition to close the description. The choice of SSC can be understood as the freedom to choose the point internal to the spinning body, whose motion along the worldline is followed, as any spinning ``particle'' must actually be an extended object, and has a lower bound for its finite size due to its spin. However, the SSC should fix the representative worldline of the body \textit{uniquely}, whereas not all choices of SSC in curved spacetime do \cite{Kyrian:2007zz}. Tulczyjew was the first to give a definition of SSC, i.e. of a center of mass, which ensures the uniqueness of the EOM \cite{Tulczyjew:1959b}.  

An appropriate SSC in curved spacetime may be written in the form 
\be\label{eq:ssc}
C^\mu \equiv S^{\mu\nu} V_\nu = 0,
\ee
where $V^\nu$ is some suitably chosen timelike vector \cite{Kyrian:2007zz}. For example, Tulczyjew suggested the covariant SSC given by \cite{Tulczyjew:1959b}
\be \label{eq:tulssc}
S^{\mu\nu} p_\nu=0, 
\ee
where $p_\nu$ is the linear momentum here also. Note that only three out of the four components of Eq.~(\ref{eq:ssc}) are independent constraints as $C^\mu V_\mu \equiv 0$. This corresponds exactly to the three redundant spin degrees of freedom that should be eliminated. Thus, the three independent constraints are taken to be the spatial components $C^i$. Also, $C^\mu$ is a zero vector, so of course the SSC can be applied in any preferred frame. In particular we may choose to apply it in the local Lorentz frame, or in the coordinate frame. The preservation of the SSC upon evolution, i.e. the requirement that $D C^\mu/D\sigma=0$, together with the MP equations Eqs.~(\ref{eq:mp1}) and (\ref{eq:mp2}), applied to the covariant SSC in Eq.~(\ref{eq:tulssc}), yields the linear momentum $p^\nu$ in terms of the coordinate velocity and the spin 
\be \label{eq:expp}
p^\nu = m\frac{u^\nu}{\sqrt{u^2}} + O(S^2), 
\ee
where the part quadratic in the particle's spin, not written explicitly, is SSC dependent. We see that if we plug $p^\nu$ from Eq.~(\ref{eq:expp}) into Eq.~(\ref{eq:lpp}), we recover the nonspinning part of the PP action given in Eq.~(\ref{eq:lppm}). 
Also if we plug Eq.~(\ref{eq:expp}) into the covariant SSC in Eq.~(\ref{eq:tulssc}), we see that in our approximation the covariant SSC can be taken as 
\be \label{eq:ssccov}
S^{\mu\nu} u_\nu=0.
\ee

Any SSC of the form of Eq.~(\ref{eq:ssc}), which guarantees the uniqueness of the solution of the EOM, can be used to eliminate the redundant degrees of freedom of spin, namely, in the end to be substituted in the temporal spin components $S_{0i}$. The basic procedure for this to be carried out is to add a Lagrange multiplier vector $\lambda_\mu$ multiplied by the SSC, i.e.~a term $\lambda_\mu C^\mu$, to the Lagrangian, as was done in e.g.~\cite{Steinhoff:2009ei}. However, the use of Lagrange multipliers to incorporate the SSC in the action would be out of place in a treatment of the problem in an EFT approach. Yee and Bander, who formally worked with a Routhian due to the explicit use of the spin momentum as an independent variable, with its conjugate coordinate being cyclic \cite{Goldstein:1980}, added an acceleration dependent piece to the Routhian given by 
\be \label{eq:yeba}
\frac{u^aS_{ab}}{u^2}\frac{Du_b}{D\sigma}, 
\ee
in order to ensure that the covariant SSC of Eq.~(\ref{eq:ssccov}) is conserved \cite{Yee:1993ya}. Note that this term is proportional to the SSC itself. Following the work of Yee and Bander a new algebraic Riemann dependent operator was also added to a Routhian in \cite{Porto:2008tb}. The additional operator given by 
\be \label{eq:sscope}
\frac{1}{2m}R_{\rho\kappa\alpha\beta} S^{\alpha\beta} S^{\rho\lambda} \frac{u^\kappa u^\lambda}{\sqrt{u^2}}
\ee
is obtained from the term in Eq.~(\ref{eq:yeba}), by using the MP equation in Eq.~(\ref{eq:mp1}), and the relation in Eq.~(\ref{eq:expp}). This term is quadratic in the spin of the particle, i.e.~of the form of a self-induced quadrupolelike operator. However, the main disadvantage of working similarly with the SSC in Eq.~(\ref{eq:ssccov}) is that it leads to noncanonical variables, i.e.~such that the EOM, in particular, the spin EOM, which are conveniently obtained from a Hamiltonian/Routhian, cannot be obtained by using the usual canonical Poisson brackets, and one will have to make some noncanonical variable transformations to arrive to canonical variables that satisfy the canonical algebra. Alternatively, the SSC can be imposed within the Hamiltonian formalism by replacing the original Poisson brackets with the so-called Dirac brackets \cite{Hanson:1974qy}. In that case the constraints in Eq.~(\ref{eq:ssc}) alone are not sufficient to describe the phase space consistently, as one needs to impose an additional set of three constraints on the configuration space variables conjugate to the spin, so that the constraint hypersurface in phase space contains the same number of configuration coordinates and conjugate momenta. The Dirac brackets are in essence the projection of the original symplectic structure onto the phase space hypersurface defined by the constraints. We note that in an action approach, there is no need for additional conjugate constraints, as was made clear in \cite{Steinhoff:2009ei} for example, where the corresponding Lagrange multipliers are found to vanish, i.e.~these constraints are redundant. Finally, the constrained Hamiltonian will be obtained simply by inserting the constraints directly in the original Hamiltonian. 

The so-called Newton-Wigner (NW) SSC is well known to be the \textit{only} SSC in flat spacetime, which yields canonical variables \cite{Hanson:1974qy,Pryce:1948pf,Newton:1949cq}. Recently, the NW SSC was generalized from flat to curved spacetime in \cite{Steinhoff:2008zr,Barausse:2009aa}. In \cite{Barausse:2009aa} they show that their generalized NW SSC leads to canonical variables at linear order in the particle's spin. A generalized NW SSC may be formulated as 
\be \label{eq:covnw}
S^{\mu\nu}(p_\nu+m e^0_\nu)=0,
\ee
where the tetrad here is the local background tetrad, and $p_\nu$ is again the linear momentum. Note that Eq.~(\ref{eq:covnw}) is of the form given in Eq.~(\ref{eq:ssc}), and that clearly Eq.~(\ref{eq:covnw}) reduces to the flat spacetime NW SSC given by \cite{Hanson:1974qy} 
\be 
S^{\mu\nu}p_\nu+mS^{\mu0}=0. 
\ee 
It should be noted however that a generalization of the NW SSC to curved spacetime is not unique. Here though, we will work with the covariant SSC of Eq.~(\ref{eq:ssccov}), which are validly incorporated into our action approach.
In the covariant SSC of Eq.~(\ref{eq:ssccov}), we have for the spin temporal components that
\be\label{eq:covst}
S^{i0} = S^{ij}v^j,
\ee
which implies a suppression of the temporal spin components $S^{i0}$ by one order of $v$ with respect to the spatial components $S^{ij}$. We note that here we apply the SSC in the local Lorentz frame, since we will work with the spin variable that is projected onto the local tetrad space, so the velocity here is the local velocity, and not the coordinate velocity. Similarly, to the order considered here in the generalized NW SSC of Eq.~(\ref{eq:covnw}) we would have   
\be\label{eq:nwst} 
S^{i0} = \frac{1}{2}S^{ij}v^j + \frac{1}{8}S^{ij}v^jv^2 + \cdots, 
\ee 
which implies a suppression of the temporal spin components $S^{i0}$ by at least one order of $v$ with respect to the spatial components $S^{ij}$, and there is an infinite expansion in the local velocity $v$.

Let us consider then the power counting of spin within the EFT description. The power counting rules in the EFT description of a binary that does not have spinning components within the PN approximation were given in \cite{Goldberger:2004jt,Goldberger:2007hy}, and we will not repeat them here. However, we should address here the power counting of a spinning compact object, in which we are concerned. We note that the spin of a compact object scales as
\be\label{eq:pcs}
S \sim mv_{rot}r_s < mr_s\sim mv^2r\sim Lv,
\ee
where $v_{rot}$ is the rotational velocity of the object, $r_s\sim Gm$ is its Schwarzschild radius, and we have used the Virial theorem $Gm/r \sim v^2$. We therefore see that the spin gets suppressed by one order of $v$ with respect to the orbital angular momentum, even for maximally rotating objects ($v_{rot}=1$), which we will take our objects to be. However, there exists another way of PN counting the spin, the so-called ``formal'' counting \cite{Steinhoff:2008zr}. In the formal counting the PN orders are counted from the original Einstein equations. Our counting adds half a PN order for each spin variable appearing in the expressions, compared to the formal counting. For example, the NLO SO and spin1-spin2 sectors are of the orders 2.5PN and 3PN, respectively, in our counting, but are both of the 2PN order in the formal counting. Our counting reflects better the true relevance of the PN spin corrections, and does not overestimate their importance. Moreover, it corresponds to the standard practice when defining the templates of LIGO/Virgo and LISA detectors. However, it does not reflect well the computational complexity involved. In the formal counting some similarities to PN calculations for nonspinning objects are more manifest. To some extent, the formal counting reflects better the computational demands, e.g.~the difficulty of the integrations required, and the regularization techniques that must be employed.   

Finally, let us focus our attention back on the linear in spin part of the PP Lagrangian given by the second term in Eq.~(\ref{eq:lpp}). It can be rewritten in terms of the Ricci rotation coefficients (also called the spin connection) defined by (up to a sign convention) \cite{Wald:1984rg}
\be
\omega_\mu^{ab}=e^{b\nu}D_\mu e^a_\nu. 
\ee
The relevant spin part of the PP Lagrangian $L_{pp(\bf{S})}$ then reads
\be
L_{pp(\bf{S})} = -\frac{1}{2}S_{ab}\Omega^{ab}-\frac{1}{2}S_{ab}\omega_\mu^{ab}u^\mu. 
\ee 
We recognize that the first term here represents the rotations of the body fixed Lorentz frame with respect to the background Lorentz frame. Hence, this kinetic term of flat spacetime, which does not contribute to the interaction of spin with gravity, will be omitted henceforth. Note that the spin that appears here is the projection of the spin tensor onto the background local tetrad, i.e.~$S^{ab} \equiv S^{\mu\nu}e^a_\mu e^b_\nu$. Therefore, the gravitational coupling to the spins of the objects is described by the relevant spin part of the PP action for each of the two objects in the binary system, so that we have for each \cite{Yee:1993ya}
\be\label{eq:lpps}
S_{pp(\bf{S})} = -\frac{1}{2} \int d\sigma S^{ab}\omega_{\mu ab}u^\mu + \cdots,
\ee
where the ellipsis denotes subleading corrections beyond linear in spin, or due to finite size effects. Now, since it is convenient to obtain the EOM for the position of the particle from a Lagrangian, whereas the EOM for the spin are easily obtained from a Hamiltonian, we may consider the spin momentum as an independent variable, where we note that its conjugate coordinate is cyclic. Thus, we can consider our Lagrangian as a Routhian for the purposes of EOM derivation, and the Legendre transformation to a Hamiltonian with respect to the spin conjugate coordinate is trivial. 

In order to derive the Feynman rules for the spin couplings, it is convenient to use the common Lorentz covariant parametrization for the expansion of the metric around flat spacetime $g_{\mu\nu} = \eta_{\mu\nu} + h_{\mu\nu}$. Then, using Eq.~(\ref{eq:gfixe}), one can find the general solution for the background tetrad, expanding in terms of $h_{\mu\nu}$
\be
e^a_\mu = \Lambda^a_\mu + \frac{1}{2}h_{\mu\nu}\Lambda^{\nu a} - \frac{1}{8}h_{\mu\rho}h^\rho_\sigma\Lambda^{\sigma a} + \cdots.
\ee 
However, we can use the internal Lorentz invariance of the background tetrad to fix our reference tetrad to be aligned with the Lorentz frame at infinity so that 
\be
e^a_\mu = \delta^a_\mu + \frac{1}{2}h^a_\mu - \frac{1}{8}h^a_\rho h^\rho_\mu + \cdots.
\ee
Using this tetrad, and expanding in Eq.~(\ref{eq:lpps}) up to second order in $h_{\mu\nu}$, which is all we will need here, we obtain the following Lagrangian: 
\be
L_{pp(\bf{S})} = \frac{1}{2}S^{ab}h_{a\mu,b} u^\mu + \frac{1}{4}S^{ab} h^\nu_b\left(\frac{1}{2}h_{a\nu,\mu}+h_{\nu\mu,a}-h_{a\mu,\nu}\right)u^\mu + \cdots.
\ee
Going back to our NRG field parametrization using Eq.~(\ref{eq:gkk}), and our choice for the worldline parameter $\sigma=t$, we obtain the following couplings to the worldline spin: 
\bea
S_{pp(\bf{S})} = \int dt &&\left[ ~\frac{1}{4} S^{ij}F_{ij}+S^{ij}\partial_j\phi v^i+S^{0i}\partial_i\phi+\frac{1}{2}S^{ij}\partial_i\sigma_{jk}v^k-\frac{1}{2}S^{0i}\partial_iA_jv^j+\frac{1}{2}S^{0i}\partial_0A_i+\cdots\right. \nn \\ 
&&\left.+S^{ij}F_{ij}\phi-\frac{1}{2}S^{ij}A_i\partial_j\phi+2S^{0i}\partial_i\phi\phi +\cdots~ \right],
\eea 
where the ellipsis stands for one-, two-, and n-graviton couplings, and velocity powers beyond the order considered here. Similarly to the mass couplings, any number of gravitons can couple to the worldline spin. Here however, unlike the couplings to the worldline mass, the order in the velocity $v$ of the couplings is not explicit, due to the $S^{0i}$ spin entries and the freedom in the choice of SSC, in addition to the possibility of having time derivatives on the fields, since the spin is derivative-coupled. Also, here it is the $A_i$ field that couples at LO in $v$, whereas the $\phi$ and $\sigma_{ij}$ fields are subleading couplings, both at the same order of $v$ (hence, the assignment ``gravito-magnetic'' for $A_i$, due to its dominant coupling to the spin). 

The Feynman rules for the one-graviton couplings to the worldline spin are thus given by
\begin{align}
\label{eq:sA}  \parbox{12mm}{\includegraphics{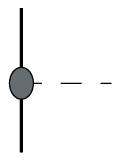}}
 & = \int dt \left[\frac{1}{2}S^{ij}(t)\partial_iA_j({\bf{x}}(t)) - \frac{1}{2}S^{0i}(t)\partial_iA_j({\bf{x}}(t))v^j(t) + \frac{1}{2}S^{0i}(t)\partial_0A_i({\bf{x}}(t))\right], \\ 
\label{eq:sphi}   \parbox{12mm}{\includegraphics{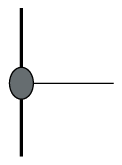}}
 & = \int dt \biggl[S^{ij}(t)\partial_j\phi({\bf{x}}(t)) v^i(t) + S^{0i}(t)\partial_i\phi({\bf{x}}(t)) + \cdots \biggr], \\  
\label{eq:ssigma}   \parbox{12mm}{\includegraphics{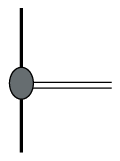}}
 & = \int dt \left[\frac{1}{2}S^{ij}(t)\partial_i\sigma_{jk}({\bf{x}}(t))v^k(t) + \cdots\right],
\end{align}
where the gray oval blobs represent the spins on the worldlines. Note the time derivative appearing in the coupling  of the $A_i$ field to the worldline spin in Eq.~(\ref{eq:sA}). This is a departure from the leading stationary approximation. Note also the $S^{0i}$ spin entry that appears at the leading order $\phi$ coupling in Eq.~(\ref{eq:sphi}). This coupling enters already at the LO spin-orbit interaction (as we will see shortly in Sec.~\ref{sec:solo}), so the use of SSC is required already at the LO gravitational spin correction at 1.5PN. Moreover, this $S^{0i}\phi$ coupling adds much complication to our calculation. We also note that the ellipsis in Eq.~(\ref{eq:sphi}) stands for another spin coupling term with a time derivative, which equals $S^{i0}\partial_0\phi\,v^i$. However, it can be seen from power counting that to the order considered here this contribution would eventually drop at the implementation of the SSC. 

For two-graviton couplings to the worldline spin the Feynman rules are 
\begin{align}
\label{eq:sphiA}  \parbox{12mm}{\includegraphics{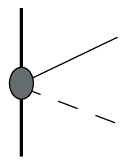}}
 & = \int dt \left[2S^{ij}(t)\partial_iA_j({\bf{x}}(t))\phi({\bf{x}}(t)) - \frac{1}{2}S^{ij}(t)A_i({\bf{x}}(t))\partial_j\phi({\bf{x}}(t)) + \cdots\right], \\ 
\label{eq:sphi^2}   \parbox{12mm}{\includegraphics{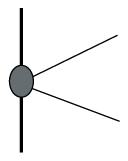}}
 & = \int dt \biggl[2S^{0i}(t) \partial_i\phi({\bf{x}}(t))\phi({\bf{x}}(t)) + \cdots \biggr].
\end{align}
Note that the two-graviton $\phi\phi$ coupling appears with the $S^{0i}$ spin entry, i.e.~the use of SSC is required here. Also recall that the velocity, appearing in the SSC derived expressions in Eqs.~(\ref{eq:covst}) and (\ref{eq:nwst}) for the $S^{i0}$ entries, is the local velocity and not the coordinate velocity, appearing here in the Feynman rules. So we actually have metric field corrections in the temporal $S^{i0}$ entries from the SSC rewritten as
\be \label{eq:sscpro}
S^{ab}e_{b\mu}u^\mu=0.
\ee 
For example, for the covariant SSC in Eq.~(\ref{eq:covst}), we have in terms of the coordinate velocity $S^{i0} = S^{ij}v^j + \frac{1}{2}S^{ij}A_{j} + \cdots$, with the gravito-magnetic vector field $A_j$ as the leading order metric field correction in the temporal spin entries, as was already found in the NLO spin1-spin2 sector \cite{Levi:2008nh}. These metric dependent corrections may be considered as contributions to the two-graviton couplings in Eqs.~(\ref{eq:sphiA}) and (\ref{eq:sphi^2}) here, arising from the one-graviton spin coupling in Eq.~(\ref{eq:sphi}) with $S^{i0}$ entries, if the SSC is applied at the level of the action.
  
It is important to note that the addition of the algebraic Riemann dependent operator from \cite{Porto:2008tb}, appearing in Eq.~(\ref{eq:sscope}), will not affect the Feynman rules relevant for the SO sector (nor for the spin1-spin2 sector), since the operator is quadratic in the spin of the particle. Finally, it should be pointed out that the fact that the spin is derivative-coupled also raises the complexity of the computations of spin corrections.

\section{Leading order spin-orbit potential}\label{sec:solo}

In this section we derive the well-known LO spin-orbit potential, originally computed in \cite{Tulczyjew:1959a,Barker:1975ae}. The result here will be important later in Sec.~\ref{sec:sonlo} in the derivation of the NLO spin-orbit Hamiltonian. Moreover, we use this most simple computation of the SO sector to illustrate the computation of Feynman diagrams that contribute to this sector, and its ambiguities related with different choices of SSC. We will show how these ambiguities can be resolved in terms of canonical and noncanonical transformations. The SO sector contains the diagrams, which have one spin coupling of one of the particles, together with at least one mass coupling of the other particle in the binary. First, we should use the power counting rules of \cite{Goldberger:2004jt,Goldberger:2007hy}, and our power counting for spin given in Eq.~(\ref{eq:pcs}), to assign a power in the velocity $v$ in the PN expansion for each term in the Feynman rules given in Secs.~\ref{sec:eftKK} and \ref{sec:spineft}. Then, we construct the SO diagrams, which contribute to a certain PN order, by considering all diagrams composed of all possible components, whose PN order adds up to the relevant PN order. The LO spin-orbit sector contains contributions only at $O(G)$, at which we have the single topology of one-graviton exchange.   

From our PN expansion of the Feynman rules, we find that the LO spin coupling is given by the first term of Eq.~(\ref{eq:sA}), i.e.~by 
\be \label{eq:ls2}
L_{({\bf{S}})2} = \frac{1}{4} S^{ij}F_{ij},
\ee
where the numerical subscript denotes the power in $v$ of this term in the PN expanded Feynman rules. This can only be contracted, using Eq.~(\ref{eq:prA}), with the LO mass coupling of $A_i$ that is given by 
\be\label{eq:lm1} 
L_{(m)1} = m A_iv^i.
\ee
\begin{figure}[b]
\includegraphics{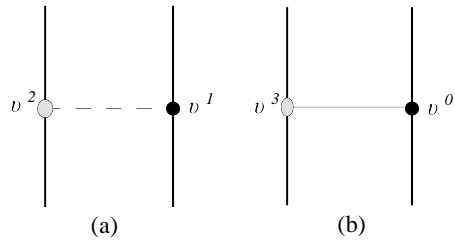}
\caption{LO spin-orbit interaction Feynman diagrams. The heavy solid lines represent the point particles worldlines. The oval gray and spherical black blobs represent the spin and mass couplings on the worldline, respectively. The solid and dashed lines represent the $\phi$ and $A_i$ fields, respectively. All diagrams must be included with their mirror images.}\label{solo}
\end{figure}
We obtain a one-graviton exchange diagram depicted in Fig.~1(a). It has the value
\be 
Fig.~1(a) = -\frac{2Gm_2}{r^2} {\bf{S}}_1\cdot{\bf{v}}_2 \times {\bf{n}} + [1 \leftrightarrow 2],
\ee
where we denote ${\bf{r}}\equiv{\bf{x}}_1(t)-{\bf{x}}_2(t)$, $r\equiv\left|\bf{r}\right|$, and ${\bf{n}}\equiv\frac{{\bf{r}}} {r}$. The spin is represented by a 3-vector defined by $S^{ij} \equiv \epsilon^{ijk}S^{k}$. The labels 1 and 2 are used for the left and right worldlines, respectively. The [$1\leftrightarrow2$] notation stands for a similar term, contributed by the mirror image of the diagram displayed, whose value is obtained under the interchange of particles labels. Note that under this exchange $\bf{n}\to\bf{-n}$. Although $\bf{r}$, ${\bf{v}}_2$, ${\bf{S}}_1$, etc.~depend on t, we suppress this dependence here and henceforth. Moreover, here and henceforth a multiplicative factor of $\int dt$ is suppressed and omitted from all diagram values. Here, we had to evaluate the Fourier integral coming from the propagator, and for that we have used Eq.~(\ref{eq:ft1}) in Appendix \ref{app:a}. It is easy to see that this diagram scales as $v^3$ (1.5PN). Hence, considering the next order in the spin couplings, we realize that there is another diagram contributing to the LO SO potential.

We go on to the spin couplings of order $v^3$ given by Eqs.~(\ref{eq:sphi}) and (\ref{eq:ssigma}), so that we have 
\be\label{eq:ls3}
L_{({\bf{S}})3} = S^{ij}\partial_j\phi v^i + S^{0i}\partial_i\phi+\frac{1}{2}S^{ij}\partial_i\sigma_{jk}v^k, 
\ee
where here the temporal spin entry $S^{0i}$ must also be taken into account, considering the SSC. However, contracting the $\sigma_{ij}$ field with the LO mass coupling of $\sigma_{ij}$ given by
\be\label{eq:lm2sigma}
L_{(m)2} = \frac{1}{2}m\sigma_{ij} v^iv^j
\ee
leads to a $v^5$ order contribution, so it cannot enter here. Thus, we contract the $\phi$ field from Eq.~(\ref{eq:ls3}), using Eq.~(\ref{eq:prphi}), with the LO mass coupling of $\phi$ given by
\be\label{eq:lm0} 
L_{(m)0} = -m \phi. 
\ee
Again, we obtain a one-graviton exchange diagram, that which is depicted in Fig.~1(b). This diagram is evaluated by
\be 
Fig.~1(b) = \frac{Gm_2}{r^2} \left[{\bf{S}}_1\cdot{\bf{v}_1}\times{\bf{n}} + S_1^{0i}n^i\right]+ [1 \leftrightarrow 2]. 
\ee
Here again, we have the Fourier integral given by Eq.~(\ref{eq:ft1}). One easily verifies this diagram also scales as $v^3$. No other diagrams can be constructed at this order. Thus, the LO spin-orbit potential is of the 1.5PN order, and it follows from these two Feynman diagrams of one-graviton exchange. We point out that both diagrams here as well as all diagrams contributing to the spin-orbit interaction are always considered together with their mirror images. Thus, the [$1\leftrightarrow2$] term will be omitted henceforth from all diagram values. 

The two diagrams here add up to the LO spin-orbit Lagrangian, which at this PN order is trivially Legendre transformed to a potential, namely $V_{SO}^{LO}=-L_{SO}^{LO}$, with the momenta conjugate to the coordinate velocity being just the Newtonian momenta, i.e.~$p=mv$. Thus, we obtain the well-known result for the LO spin-orbit potential 
\be\label{eq:vlo}
V_{SO}^{LO} = -\frac{Gm_2}{r^2} \left[{\bf{S}}_1\cdot\left({\bf{v}}_1\times{\bf{n}} - 2{\bf{v}}_2\times{\bf{n}}\right) + S_1^{0i}n^i\right] + [1 \leftrightarrow 2]
= -\frac{Gm_2}{r^2} {\bf{S}}_1\cdot \left(\frac{3}{2}{\bf{v}}_1\times{\bf{n}}-2{\bf{v}}_2\times\bf{n}\right) + [1 \leftrightarrow 2],
\ee
where on the right-hand side the leading term of NW SSC from Eq.~(\ref{eq:nwst}) is substituted in, and the result is then in agreement with the canonical result found in e.g.~Eq.~(4.11a) of \cite{Damour:2007nc}. We would like to stress that unlike the LO spin1-spin2 potential, the LO spin-orbit potential already requires the use of SSC, a fact which demands our special attention. It is the $S^{0i}$ spin entry that appears at the leading $\phi$ spin coupling which adds complication to our calculation. 

Let us consider then the LO spin-orbit potential given in Eq.~(\ref{eq:vlo}) with the use of the covariant SSC of Eq.~(\ref{eq:ssccov}), instead of the NW SSC. Upon the use of the covariant SSC from Eq.~(\ref{eq:covst}) in Eq.~(\ref{eq:vlo}), the LO spin-orbit potential takes the form 
\be\label{eq:vlocov}
V_{SO}^{LO} = -\frac{Gm_2}{r^2} {\bf{S}}_1\cdot \left(2{\bf{v}}_1\times{\bf{n}}-2{\bf{v}}_2\times\bf{n}\right) + [1 \leftrightarrow 2]= -2\frac{Gm_2}{r^2} {\bf{S}}_1\cdot {\bf{v}} \times{\bf{n}} + [1 \leftrightarrow 2],
\ee
where in the last equality we have used the relative velocity ${\bf{v}}\equiv{\bf{v}}_1-{\bf{v}}_2$. However, the known canonical result for the LO spin-orbit Lagrangian obtained with the use of covariant SSC takes the form \cite{Damour:1982}
\be\label{eq:vlocovcan}
L_{SO}^{LO} = 2\frac{Gm_2}{r^2} {\bf{S}}_1\cdot {\bf{v}} \times{\bf{n}} + \frac{1}{2}{\bf{S}}_1\cdot {\bf{v}}_1 \times{\dot{\bf{v}}}_1 + [1 \leftrightarrow 2],
\ee
where ${\dot{\bf{v}}_a}\equiv d {\bf{v}}_a/dt$ is the acceleration of the particle. It is easy to see that Eq.~(\ref{eq:vlocovcan}) differs from the result obtained via the EFT calculation in Eq.~(\ref{eq:vlocov}) by an additional acceleration dependent piece. One may be reluctant to have such terms, since they call for the use of the EOM at the level of the action, a procedure which is known to be incorrect in many (in)famous examples. However, as originally noted by \cite{Schafer:1984}, and also treated by many others, e.g.~\cite{Barker:1980,Damour:1990jh}, a substitution of low order EOM in higher order terms in the level of the Lagrangian is a correct procedure in GR, and is equivalent to making a coordinate transformation. Here, we may eliminate the acceleration terms using the LO EOM given by 
\be\label{eq:aeom}
{\dot{\bf{v}}}\equiv{\bf{a}}={\bf{a}}_1-{\bf{a}}_2=-\frac{G(m_1+m_2)}{r^2}{\bf{n}},  
\ee
and we see that the canonical result obtained with the NW SSC in Eq.~(\ref{eq:vlo}) is recovered. 

Alternatively then, one can recognize that the canonical result obtained with the covariant SSC in Eq.~(\ref{eq:vlocovcan}), is related to the canonical result obtained with the NW SSC in Eq.~(\ref{eq:vlo}), by the following coordinate transformation at the level of the Lagrangian 
\be
{\bf{r}}_a\to {\bf{r}}_a+\frac{1}{2m_a}{\bf{S}}_a\times {\bf{v}}_a \,\,\,\, \Rightarrow \,\,\,\,
{\bf{v}}_a\to {\bf{v}}_a+\frac{1}{2m_a}{\bf{S}}_a\times {\dot{\bf{v}}}_a,
\ee
where the spin precession ${\dot{\bf{S}}}$ is considered at Newtonian order as 0. Equivalently, they are related by the following \textit{canonical} transformation at the level of the Hamiltonian
\bea
{\bf{r}}_a & \to & {\bf{r}}_a+\frac{1}{2m_a^2}{\bf{S}}_a\times {\bf{p}}_a, \\
{\bf{p}}_a & \to & {\bf{p}}_a-\frac{1}{2m_a}{\bf{S}}_a\times {\bf{\dot{p}}}_a.
\eea
So now it becomes clear how to relate the EFT result obtained with the covariant SSC, to the canonical results obtained with the use of the covariant SSC or NW SSC. To relate Eq.~(\ref{eq:vlocov}) with Eq.~(\ref{eq:vlocovcan}) one should make the following \textit{noncanonical} change of variables at the level of the Hamiltonian:
\bea
{\bf{p}}_a & \to & {\bf{p}}_a+\frac{1}{2m_a}{\bf{S}}_a\times {\bf{\dot{p}}}_a,
\eea
transforming only p, and not r, whereas to relate Eq.~(\ref{eq:vlocov}) with Eq.~(\ref{eq:vlo}) one should make the following \textit{noncanonical} change of variables at the level of the Hamiltonian:
\bea \label{eq:covtonw}
{\bf{r}}_a & \to & {\bf{r}}_a+\frac{1}{2m_a^2}{\bf{S}}_a\times {\bf{p}}_a,
\eea
transforming only r, and not p. The transformation in Eq.~(\ref{eq:covtonw}) is just similar to the flat spacetime mapping from the center of mass coordinate variable related with the covariant SSC, to that related with NW SSC \cite{Hanson:1974qy}. To conclude, we have seen that already at the LO of the spin-orbit interaction the result obtained with an EFT calculation using the covariant SSC leads to a noncanonical result, related to the canonical results by noncanonical change of variables. 

\section{Next-to-leading order spin-orbit interaction} \label{sec:csonlo}

In this section, we present the calculation of the Feynman diagrams, which contribute to the NLO spin-orbit interaction potential between the binary constituents. In the NLO spin-orbit potential, which is of the 2.5PN order, we have $O(G)$ and $O(G^2)$ contributions, so there are three diagram topologies contributing. First, we have the single topology at $O(G)$ of one-graviton exchange, and then we have the two topologies at $O(G^2)$ \cite{Gilmore:2008gq}. At $O(G^2)$ we first consider the ``V'' topology, i.e.~diagrams of two-graviton exchange. Then, we must consider the ``Y'' topology, namely, diagrams that contain a three-graviton vertex. Note that after stripping off worldlines from our diagrams, we are always left with connected diagrams, and no graviton loops. We use the Feynman rules presented in Secs.~\ref{sec:eftKK} and \ref{sec:spineft} for the propagators, propagator correction vertices, three-graviton vertices, and the worldline mass and spin couplings. In addition, we have to keep track in each diagram of the possible numerous contractions, and the symmetry factors arising from the worldline couplings at $O(G^2)$ \cite{Peskin:1995ev}. We evaluate each diagram in turn, starting from the simplest one-graviton exchange diagrams and ending with the more complex time dependent cubic self-gravitational interaction diagrams.

\subsection{One-graviton exchange}

For the NLO of the spin-orbit interaction, we start by still considering one-graviton exchange diagrams. Calculating now to order $v^5$ (2.5PN), there are seven diagrams to be evaluated as shown in Fig.~2.

We should include diagrams with a spin coupling of $L_{(\bf{S})2}$ from Eq.~(\ref{eq:ls2}), and the subleading mass coupling from Eq.~(\ref{eq:mA}) given by
\be\label{eq:lm3}
L_{m(3)} = \frac{1}{2}m A_iv^i v^2.
\ee
We obtain the diagram shown in Fig.~2(a1). The value of this diagram is given by
\be 
Fig.~2(a1) = -\frac{Gm_2}{r^2} {\bf{S}}_1\cdot{\bf{v}}_2\times{\bf{n}}~v_2^2.
\ee
The only integral in $\bf{k}$ which arises is a Fourier transform, evaluated using the formula in Eq.~(\ref{eq:ft1}). This is the case for all one-graviton exchange diagrams that do not include propagator corrections, i.e.~for Figs.~2(a1)-2(a4). 

For the NLO of the SO interaction, we should also include diagrams with a spin coupling of $L_{(\bf{S})3}$ in Eq.~(\ref{eq:ls3}) and the  mass couplings of $\phi$ given by
\be\label{eq:lm2}
L_{(m)2} = -\frac{3}{2}m \phi v^2,
\ee
and of $\sigma_{ij}$ from Eq.~(\ref{eq:lm2sigma}). Contracting each of the fields $\phi$ and $\sigma_{ij}$, we obtain the diagrams shown in Figs.~2(a2.1) and 2(a2.2), respectively. The values of these diagrams are given by
\bea 
Fig.~2(a2.1) &=& \frac{3Gm_2}{2r^2} \left[{\bf{S}}_1\cdot{\bf{v}}_1\times{\bf{n}} + S_1^{0i}n^i\right]v_2^2,
\label{eq:2a2.1}\\
Fig.~2(a2.2) &=& -\frac{2Gm_2}{r^2} {\bf{S}}_1\cdot \left[ {\bf{v}}_1\times{\bf{n}}~v_2^2-{\bf{v}}_2\times{\bf{n}}~({\bf{v}}_1\cdot{\bf{v}}_2) \right].
\eea
\begin{figure}[b]
\includegraphics{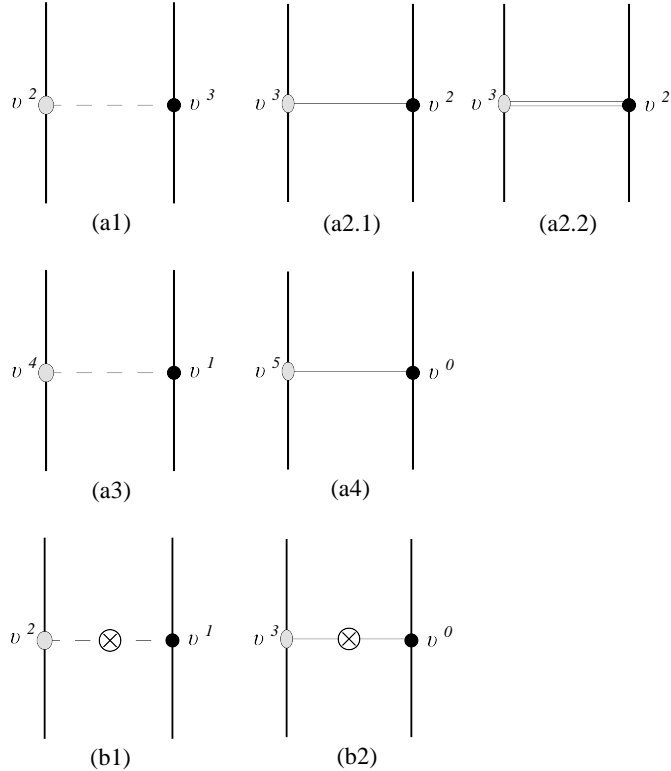}
\caption{NLO spin-orbit interaction Feynman diagrams of one-graviton exchange. The double line represents the $\sigma_{ij}$ field. The encircled cross vertex corresponds to a propagator correction. All diagrams should be included with their mirror images.}\label{solnlo}
\end{figure}

Next, we must consider the spin coupling from Eq.~(\ref{eq:sA}) at order $v^4$ given by 
\be\label{eq:ls4}
L_{({\bf{S}})4} = -\frac{1}{2}S^{0i}\partial_iA_jv^j + \frac{1}{2}S^{0i}\partial_0A_i. 
\ee
Note that both terms here contain temporal spin entries. Moreover, note the second term here that first contains time dependence. We should then also include diagrams with the spin coupling of $L_{(\bf{S})4}$ together with the mass coupling $L_{(m)1}$ of Eq.~(\ref{eq:lm1}). Such a diagram is depicted in Fig.~2(a3) and is evaluated by
\be\label{eq:cd}
Fig.~2(a3) = -\frac{2Gm_2}{r^2}\left[S_1^{0i}n^i({\bf{v}}_1\cdot{\bf{v}}_2) + S_1^{0i}v_2^i({\bf{v}}_1\cdot{\bf{n}})\right] + \frac{2Gm_2}{r}\partial_tS_1^{0i}v_2^i.    
\ee 
Here, when contracting the $\partial_0A_i$ term of Eq.~(\ref{eq:ls4}), we make an integration by parts to drop the time derivative from the delta function on the Fourier transform integral, and the temporal spin component $S_{0i}$. Only then do we perform the integration over time, and use the delta function. Note that the time derivative of the temporal spin component may yield an acceleration dependent term from the implicit velocity power it carries, see e.g.~the SSC in Eq.~(\ref{eq:covst}), as well as a spin precession term. Here, we may eliminate the acceleration terms using the LO EOM given in Eq.~(\ref{eq:aeom}), which will result in a nonlinear $O(G^2)$ contribution. By such a substitution we will implicitly have a coordinate transformation made. As noted, the time derivative may also yield a spin precession term that requires the use of the LO spin EOM. However, the LO spin EOMs scale as \cite{Faye:2006gx}
\be\label{eq:spineom}
\frac{dS}{dt}\sim \frac{Gm}{r^2} Sv.
\ee
This will yield a subleading order contribution and should therefore be dropped here \cite{Levi:2008nh}. Because of the term with the time derivative this diagram may be evaluated in two ways. The contraction with the $\partial_0A_i$ term may be evaluated alternatively by flipping the time derivative between the two particles, namely, by using the identity
\be\label{eq:timeflip}
\int dt_1dt_2~\partial_{t_1}\delta(t_1-t_2)f(t_1)g(t_2)=-\int dt_1dt_2~\partial_{t_2}\delta(t_1-t_2)f(t_1)g(t_2).
\ee
Evaluating the diagram this way yields the following seemingly different value for the diagram in Fig.~2(a3), given by
\be
Fig.~2(a3)_{alt.} = -\frac{2Gm_2}{r^2}\left[S_1^{0i}n^i({\bf{v}}_1\cdot{\bf{v}}_2) + S_1^{0i}v_2^i({\bf{v}}_2\cdot{\bf{n}})\right] - \frac{2Gm_2}{r}S_1^{0i}a_2^i. 
\ee
Hence, we conclude that an acceleration dependent term and the use of EOM are inevitably involved in evaluating this diagram (though we may avoid a spin precession term on the later evaluation). We stress that the two evaluations can be related by a total time derivative, and are thus physically equivalent. 

Next, considering spin couplings at order $v^5$, we take into account the flat spacetime subleading substitution in the SSC of Eq.~(\ref{eq:nwst}). It may make the $S^{0i}$ term of Eq.~(\ref{eq:sphi}) become a spin coupling of order $v^5$ given by
\be\label{eq:ls5}
L_{({\bf{S}})5} = S^{0i}\partial_i\phi. 
\ee
Contracting this spin coupling, using Eq.~(\ref{eq:prsigma}), with the LO mass coupling of Eq.~(\ref{eq:lm0}), we may get the diagram shown in Fig.~2(a4), which is evaluated by
\be
Fig.~2(a4) = \frac{Gm_2}{r^2} S_1^{0i}n^i. 
\ee 
We should stress that we do not have this contribution, if we use the covariant SSC at the level of the action. 

There are further contributions to the NLO SO interaction from the one-graviton exchange sector, which arise from corrections to the potential graviton propagators. These are the self-gravitational vertices of Eq.~(\ref{eq:2gt}), containing two time derivatives, and hence suppressed by $v^2$. First, we consider diagrams that contain the LO spin coupling of Eq.~(\ref{eq:ls2}), contracted with the propagator correction in Eq.~(\ref{eq:prtA}), and the mass coupling of Eq.~(\ref{eq:lm1}), using Eq.~(\ref{eq:prA}). Such a diagram, which is shown in Fig.~2(b1), is evaluated by
\be
Fig.~2(b1) = \frac{Gm_2}{r^2}{\bf{S}}_1\cdot \left[-{\bf{v}}_2\times{\bf{n}}~({\bf{v}}_1\cdot{\bf{v}}_2) + {\bf{v}}_1\times{\bf{v}}_2~({\bf{v}}_2\cdot{\bf{n}}) + 3{\bf{v}}_2\times{\bf{n}}~({\bf{v}}_1\cdot{\bf{n}})({\bf{v}}_2\cdot{\bf{n}}) \right] + \frac{Gm_2}{r}{\bf{S}}_1\cdot {\bf{v}}_1\times{\bf{a}}_2.  
\ee 
Note that here we have to consider two similar contractions. To compute this diagram we first have to use the identity of Eq.~(\ref{eq:timeflip}) twice -- for each time derivative, so that we get $\partial_{t_1}$ and $\partial_{t_2}$. Then, we make two integrations by parts to drop the time derivatives on the respective Fourier integrals, or on $S_1^{ij}$ or $v_2$, respectively. Then, we apply the time derivatives. Thus, we get a spin precession term that drops on the use of the LO spin EOM from Eq.~(\ref{eq:spineom}), and an acceleration dependent term with $a_2$, to be substituted using Eq.~(\ref{eq:aeom}), as explained above. Only after using the time derivatives do we perform the time integrations, using the delta functions. Finally, we arrive at a Fourier integral, whose value is given in Eq.~(\ref{eq:ft2}). 

As for the diagrams including the propagator correction of Eq.~(\ref{eq:prtphi}), they contain the spin coupling of $\phi$ in Eq.~(\ref{eq:ls3}), and the mass coupling of Eq.~(\ref{eq:lm0}), all contracted using Eq.~(\ref{eq:prphi}). Such a diagram, shown in Fig.~2(b2), is evaluated by 
\bea 
Fig.~2(b2) &&= \frac{Gm_2}{2r^2}{\bf{S}}_1\cdot \left[{\bf{v}}_1\times{\bf{n}}~({\bf{v}}_1\cdot{\bf{v}}_2) + {\bf{v}}_1\times{\bf{v}}_2~({\bf{v}}_1\cdot{\bf{n}}) - 3{\bf{v}}_1\times{\bf{n}}~({\bf{v}}_1\cdot{\bf{n}})({\bf{v}}_2\cdot{\bf{n}}) \right] + \frac{Gm_2}{2r}{\bf{S}}_1\cdot {\bf{v}}_2\times {\bf{a}}_1 \nn \\ 
&&+ \frac{Gm_2}{2r^2}\left[S_1^{0i}n^i({\bf{v}}_1\cdot{\bf{v}}_2) + S_1^{0i}v_2^i({\bf{v}}_1\cdot{\bf{n}}) - 3S_1^{0i}n^i({\bf{v}}_1\cdot{\bf{n}})({\bf{v}}_2\cdot{\bf{n}}) \right] - \frac{Gm_2}{2r}\partial_{t}S_1^{0i}v_2^i. 
\eea 
Here also we have two similar contractions. The evaluation is similar to that of the previous diagram, except that the time derivative falls on a temporal spin component $S_1^{0i}$, which may yield an additional acceleration term upon the use of SSC, e.g.~in Eq.~(\ref{eq:covst}), as in the evaluation of Fig.~2(a3). In addition, we get terms containing $S_1^{0i}v_1^i$ and $\partial_t S_1^{0i}n^i$, which can be seen from power counting to eventually drop at the implementation of the SSC. 
Here again, acceleration dependent terms and the use of EOM are inevitable.

\subsection{Two-graviton exchange}

At NLO in the SO interaction nonlinear contributions must also be included. At $O(G^2)$, we consider first the two-graviton exchange, i.e.~the diagrams of the V topology. There are four such diagrams to be evaluated here as shown in Fig.~3. They arise from two-graviton couplings of both mass and spin. First, we consider the two-graviton mass couplings, which are simpler. At order $v^3$ we have the two-graviton mass coupling from Eq.~(\ref{eq:mphiA}), i.e.
\be\label{lmg3}
L_{(m)3} = m\phi A_iv^i.
\ee
This must be contracted with the LO spin coupling of Eq.~(\ref{eq:ls2}), using Eq.~(\ref{eq:prA}), and the LO mass coupling of Eq.~(\ref{eq:lm0}), using Eq.~(\ref{eq:prphi}), to obtain the diagram depicted in Fig.~3(a1), which equals
\be
Fig.~3(a1) = \frac{2G^2m_1m_2}{r^3}~{\bf{S}}_1\cdot{\bf{v}}_2\times{\bf{n}}.
\ee 
The calculation of the two-graviton exchange diagrams is rather simple since the two momentum integrations just factorize into the two simple Fourier transforms of the one-graviton exchange, i.e.~that of Eq.~(\ref{eq:ft1}). 
\begin{figure}[t]
\includegraphics{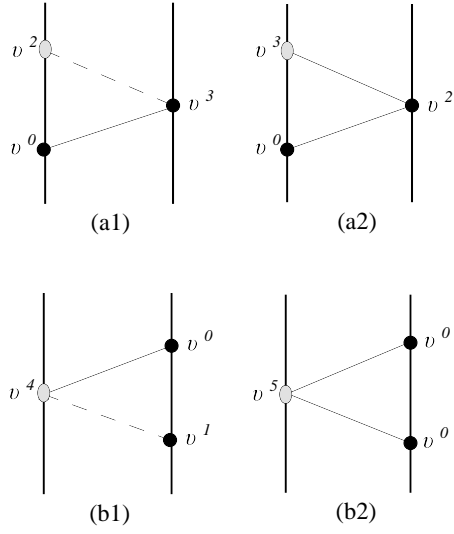}
\caption{Nonlinear NLO spin-orbit interaction Feynman diagrams of two-graviton exchange. These diagrams should be included together with their mirror images.}\label{so2nlnlo}
\end{figure}

We must also take into account the two-graviton mass coupling of Eq.~(\ref{eq:mphi^2}) given by
\be\label{lmg2}
L_{(m)2} = -\frac{m}{2}\phi^2.
\ee 
This should be contracted with the spin coupling of $\phi$ in Eq.~(\ref{eq:ls3}), and LO mass coupling of Eq.~(\ref{eq:lm0}), using Eq.~(\ref{eq:prphi}), to obtain the diagram depicted in Fig.~3(a2), which equals
\be
Fig.~3(a2) = -\frac{G^2m_1m_2}{r^3}\left[{\bf{S}}_1\cdot{\bf{v}}_1\times{\bf{n}}+S_1^{0i}n^i\right]. 
\ee 
Note that here we have two similar contractions. 

Next, we should consider the two-graviton spin couplings. First, we have from Eq.~(\ref{eq:sphiA}) the LO two-graviton spin coupling given by 
\be\label{eq:lsg4} 
L_{({\bf{S}})4}= S^{ij}F_{ij}\phi - \frac{1}{2}S^{ij}A_i\partial_j\phi. 
\ee 
The spin coupling of Eq.~(\ref{eq:lsg4}) scales as $v^4$, so diagrams containing it must be contracted with the mass couplings of Eqs.~(\ref{eq:lm0}) and (\ref{eq:lm1}), using Eqs.~(\ref{eq:prphi}) and (\ref{eq:prA}). Such a diagram is shown in Fig.~3(b1), and it equals 
\be \label{eq:s0ifield1}
Fig.~3(b1) = \frac{10G^2m_2^2}{r^3}~{\bf{S}}_1\cdot{\bf{v}}_2\times{\bf{n}}. 
\ee  

In addition, we have a contribution from the $O(v^5)$ two-graviton spin coupling of Eq.~(\ref{eq:sphi^2}), given by 
\be\label{eq:lsg5}
L_{({\bf{S}})5} = 2S^{0i}\partial_i\phi\phi. 
\ee 
Note that this two-graviton spin coupling appears with the $S^{0i}$ entry, so the use of SSC is involved. This two-graviton spin coupling should be contracted with two of the mass couplings of Eq.~(\ref{eq:lm0}), using Eq.~(\ref{eq:prphi}). This forms the diagram depicted in Fig.~3(b2) and given by the value
\be \label{eq:s0ifield2}
Fig.~3(b2) = \frac{2G^2m_2^2}{r^3}~S_1^{i0}n^i.  
\ee  
Here, the symmetry factor is $\frac{1}{2}$ but we have two similar contractions.

Recall that there are metric field corrections in the temporal $S^{i0}$ entries from Eq.~(\ref{eq:sscpro}). These may be considered as additional contributions to the two-graviton spin couplings in Eqs.~(\ref{eq:lsg4}) and (\ref{eq:lsg5}) here, arising from the $S^{0i}\partial_i\phi$ coupling in Eq.~(\ref{eq:sphi}), if the SSC is applied at the level of the action. There is an additional contribution to the $SA\phi$ coupling in Eq.~(\ref{eq:lsg4}), which was already found in the NLO spin1-spin2 sector \cite{Levi:2008nh}, and to the $S\phi^2$ coupling in Eq.~(\ref{eq:lsg5}), which did not appear in the NLO spin1-spin2 sector. We will consider these additional contributions in Sec.~\ref{sec:sonlo}.

\subsection{Cubic self-gravitational interaction}
 
Finally, we must include the nonlinear contribution arising from diagrams, where a three-graviton vertex mediates the interaction, i.e.~the second topology at $O(G^2)$ -- the Y topology \cite{Gilmore:2008gq}. As shown in Fig.~4, there are seven such diagrams to evaluate in the NLO SO interaction. All of these diagrams contain one-loop integrals and require regularization. The regularization can be understood to arise from the fact that unlike n-graviton exchange between two worldlines, in which the graviton momenta have a maximal cutoff, set by the reciprocal of the separation distance $r$ between the two worldlines, here it is possible for the three-graviton vertex to be arbitrarily close to the worldlines (in position space, of course), thereby yielding arbitrarily large momenta for the gravitons in the cubic interaction. It may also be interpreted so that these diagrams have a dressed vertex, namely, that the two vertices of the same object can be replaced by a single dressed vertex \cite{Kol:2009mj}. The loop integrals are handled with dimensional regularization by the usual techniques, see e.g.~\cite{Collins:1984xc}, and Appendix \ref{app:a} for the required formulas. It should be stressed that the NRG field parametrization we employ here certainly makes the computation of the cubic gravitational interaction diagrams feasible by hand, compared with the long calculation required in e.g.~\cite{Goldberger:2004jt} or \cite{Porto:2008tb}, which relies on the aid of automated computations. 
\begin{figure}[b]
\includegraphics{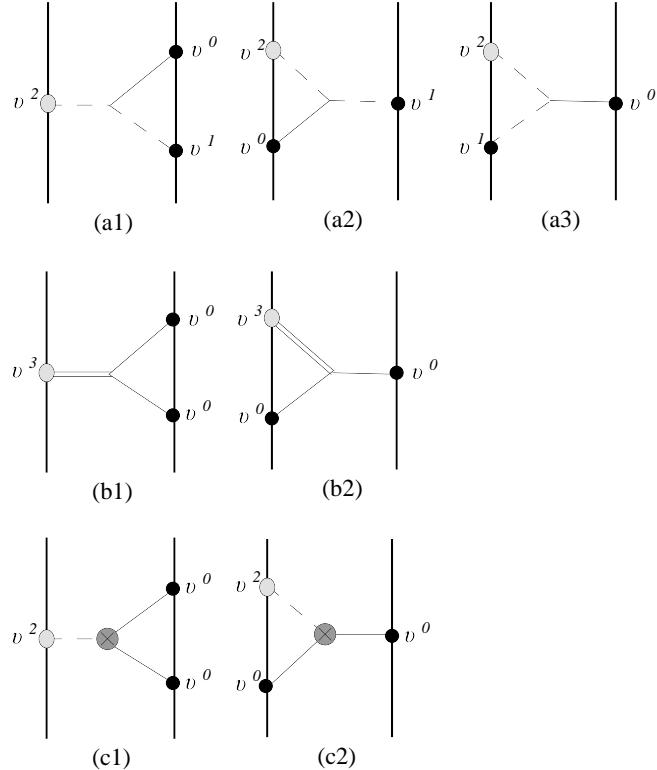}
\caption{Nonlinear NLO spin-orbit interaction Feynman diagrams with a three-graviton vertex. The gray encircled cross vertex corresponds to the time dependent cubic vertex. These diagrams should be included together with their mirror images.}\label{so3nlnlo}
\end{figure} 

The LO three-graviton vertices of Eq.~(\ref{eq:3kk}) scale as $v^2$ \cite{Goldberger:2004jt}. First, we consider the $\phi F^2$ vertex of Eq.~(\ref{eq:phiA^2}). This vertex should be contracted with the LO spin coupling of Eq.~(\ref{eq:ls2}), and the mass couplings of Eqs.~(\ref{eq:lm0}) and (\ref{eq:lm1}), using Eqs.~(\ref{eq:prphi}) and (\ref{eq:prA}) respectively. However, one should consider all possible insertions of the mass couplings on the two worldlines. One can have both mass couplings on the same worldline as shown in Fig.~4(a1), or each mass coupling on another worldline, where in the later possibility an additional permutation can be made, thus yielding the two diagrams appearing in Figs.~4(a2) and 4(a3). These diagrams are evaluated, respectively, as
\bea
Fig.~4(a1) &=& -\frac{8G^2m_2^2}{r^3}~{\bf{S}}_1\cdot{\bf{v}}_2\times{\bf{n}},\\
Fig.~4(a2) &=& -\frac{2G^2m_1m_2}{r^3}~{\bf{S}}_1\cdot{\bf{v}}_2\times{\bf{n}},\\ 
Fig.~4(a3) &=& \frac{2G^2m_1m_2}{r^3}~{\bf{S}}_1\cdot{\bf{v}}_1\times{\bf{n}}. 
\eea
In each of these diagrams, we have two similar contractions. In all of them we have to evaluate a momentum integral over two momenta. To calculate it, we first perform the integration that corresponds to a one-loop integral and then handle the remaining Fourier transform integral over the other momentum. In Fig.~4(a1), we use the one-loop integral given in Eq.~(\ref{eq:1l1}), and then we should use the Fourier integral of Eq.~(\ref{eq:ft0.5}). We note also that in Fig.~4(a1) the contraction of the second term in the vertex of Eq.~(\ref{eq:phiA^2}) cancels out, due to the antisymmetry of the spin tensor. In Figs.~4(a2) and 4(a3), we use the one-loop integral given in Eq.~(\ref{eq:1l2}) and again the Fourier integral of Eq.~(\ref{eq:ft0.5}).

Next, we also have the $\sigma\phi^2$ vertex given in Eq.~(\ref{eq:sigmaphi^2}). This vertex is contracted with the $\sigma$ term of the spin coupling in Eq.~(\ref{eq:ls3}), using Eq.~(\ref{eq:prsigma}), and with two of the mass couplings in Eq.~(\ref{eq:lm0}), using Eq.~(\ref{eq:prphi}). Here also, one can either have both mass couplings on the same worldline, leading to the diagram shown in Fig.~4(b1), or each mass coupling on another worldline, leading to the diagram shown in Fig.~4(b2). These diagrams are evaluated, respectively, as
\bea
Fig.~4(b1) &=& -\frac{G^2m_2^2}{2r^3}~{\bf{S}}_1\cdot{\bf{v}}_1\times{\bf{n}}, \\
Fig.~4(b2) &=& -\frac{G^2m_1m_2}{2r^3}~{\bf{S}}_1\cdot{\bf{v}}_1\times{\bf{n}}.
\eea
The symmetry factor of Fig.~4(b1) is $\frac{1}{2}$. Both diagrams have two similar contractions. In both diagrams one is required to use both one-loop integrals given in Eqs.~(\ref{eq:1l1}) and (\ref{eq:1l2}), and the Fourier integrals given in Eqs.~(\ref{eq:ft0.5}) and (\ref{eq:ft1.5}).

At the NLO SO interaction we have in addition a contribution from another three-graviton vertex that contains a single time derivative. This is the $A \phi^2$ vertex given in Eq.~(\ref{eq:Aphi^2t}), which scales as $v^3$. This vertex is contracted with the LO spin coupling of Eq.~(\ref{eq:ls2}), using Eq.~(\ref{eq:prA}), and with two of the LO mass couplings of Eq.~(\ref{eq:lm0}), using Eq.~(\ref{eq:prphi}). Here again, just like the diagrams that include the $\sigma\phi^2$ vertex shown in Fig.~4(b), one can either have both mass couplings on the same worldline, leading to the diagram shown in Fig.~4(c1), or each mass coupling on another worldline, leading to the diagram shown in Fig.~4(c2). These diagrams are evaluated, respectively, as
\bea
Fig.~4(c1) &=& \frac{G^2m_2^2}{2r^3}~{\bf{S}}_1\cdot{\bf{v}}_2\times{\bf{n}},\\ 
Fig.~4(c2) &=& \frac{G^2m_1m_2}{2r^3}~{\bf{S}}_1\cdot{\bf{v}}_1\times{\bf{n}}.
\eea
Here, just like with the propagator corrections in Figs.~2(b1) and 2(b2), we have the further complication of the time derivative on the gravitational vertex. Just like we did there, we first use the identity of Eq.~(\ref{eq:timeflip}) and then integrate by parts to drop the time derivatives on the Fourier integrals. Subsequently, we apply the time derivatives and only then do we perform the time integrations, using the delta functions. In both diagrams one uses both one-loop integrals given in Eqs.~(\ref{eq:1l1}) and (\ref{eq:1l2}), and the Fourier integrals given in Eqs.~(\ref{eq:ft0.5}) and (\ref{eq:ft1.5}). Figure 4(c1) has a symmetry factor of $\frac{1}{2}$ and two similar contractions. Figure 4(c2) has two different contractions, where the second contraction, in which one contracts the time derivative with the mass coupling of particle 2, vanishes, due to the antisymmetry of the spin tensor.

We note that in the NRG field parametrization used here the cubic gravitational interaction sector does not contain temporal spin entries, which is advantageous in this more computationally complex sector. Finally, it is also worth noting that had we used the fully harmonic gauge rather than the linearized harmonic gauge, which would affect the three-graviton vertices used here, we would still obtain the same values for the cubic self-gravitational interaction diagrams. Hence, we can conclude that the total NLO spin-orbit interaction result that we get would be unchanged. 

\section{Next-to-leading order spin-orbit potential}\label{sec:sonlo}

Summing up all of the contributions from the Feynman diagrams, we obtain the NLO spin-orbit interaction Lagrangian. As was noted already, we can consider our Lagrangian as a Routhian, since the Legendre transformation to a Routhian with respect to the spin conjugate coordinate is trivial. Here, we present only the ${\bf{S}}_1$ dependent part, and one should add a similar term for ${\bf{S}}_2$ by exchanging particle labels 1 and 2 in the following expression. Our NLO spin-orbit interaction Lagrangian for a binary system is then given by
\bea\label{eq:so}
L_{SO}^{NLO} = \frac{Gm_2}{r^2}{\bf{S}}_1\cdot &&
\left[{\bf{v}}_1\times{\bf{n}}\left(
\frac{1}{2}{\bf{v}}_1\cdot{\bf{v}}_2-\frac{1}{2}v_2^2-\frac{3}{2}({\bf{v}}_1\cdot{\bf{n}})({\bf{v}}_2\cdot{\bf{n}})\right) \right. \nn \\
&&+ 
\left.{\bf{v}}_2\times{\bf{n}}\biggl({\bf{v}}_1\cdot{\bf{v}}_2-v_2^2+3({\bf{v}}_1\cdot{\bf{n}})({\bf{v}}_2\cdot{\bf{n}})\biggr) +
{\bf{v}}_1\times{\bf{v}}_2\left(\frac{1}{2}{\bf{v}}_1\cdot{\bf{n}}+{\bf{v}}_2\cdot{\bf{n}}\right)\right] \nn \\
+\frac{Gm_2}{r}{\bf{S}}_1\cdot &&
\left[{\bf{v}}_1\times{\bf{a}}_2+\frac{1}{2}{\bf{v}}_2\times{\bf{a}}_1\right]+\frac{G^2m_2}{r^3}{\bf{S}}_1\cdot
\left[{\bf{v}}_1\times{\bf{n}}\left(m_1-\frac{1}{2}m_2\right)+{\bf{v}}_2\times{\bf{n}}\left(\frac{5}{2}m_2\right)\right] \nn \\
+ \frac{Gm_2}{r^2}&&
\left[
S_1^{0i}n^i\left(
1-\frac{3}{2}{\bf{v}}_1\cdot{\bf{v}}_2+\frac{3}{2}v_2^2-\frac{3}{2}({\bf{v}}_1\cdot{\bf{n}})({\bf{v}}_2\cdot{\bf{n}})\right) + S_1^{0i}v_2^i\left(-\frac{3}{2}{\bf{v}}_1\cdot{\bf{n}}\right)\right] \nn \\
+ \frac{Gm_2}{r}&&
\left[\frac{3}{2}\partial_tS_1^{0i}\cdot v_2^i\right] - \frac{G^2m_2}{r^3}
S_1^{0i}n^i \left[m_1+2m_2 
\right]. 
\eea
Here, we give the result with the acceleration dependent terms, and we do not substitute the SSC in the $S^{0i}$ dependent terms, so that the last two lines of Eq.~(\ref{eq:so}) represent the SSC dependent part. Naturally, the additional contributions of the field corrections in the $S^{0i}$ entries from Eq.~(\ref{eq:sscpro}), which may be considered in the two-graviton exchange sector, if the SSC is considered on the level of the action, are also not included in this form. 

First, we substitute in Eq.~(\ref{eq:so}) the accelerations using the LO EOM given in Eq.~(\ref{eq:aeom}) as was discussed already. We obtain
\bea\label{eq:nossc}
L_{SO}^{NLO} = \frac{Gm_2}{r^2}{\bf{S}}_1\cdot &&
\left[{\bf{v}}_1\times{\bf{n}}\left(
\frac{1}{2}{\bf{v}}_1\cdot{\bf{v}}_2-\frac{1}{2}v_2^2-\frac{3}{2}({\bf{v}}_1\cdot{\bf{n}})({\bf{v}}_2\cdot{\bf{n}})\right) \right.\nn\\
&&+ 
\left.{\bf{v}}_2\times{\bf{n}}\biggl({\bf{v}}_1\cdot{\bf{v}}_2-v_2^2+3({\bf{v}}_1\cdot{\bf{n}})({\bf{v}}_2\cdot{\bf{n}})\biggr) +
{\bf{v}}_1\times{\bf{v}}_2\left(\frac{1}{2}{\bf{v}}_1\cdot{\bf{n}}+{\bf{v}}_2\cdot{\bf{n}}\right)\right]\nn\\
+\frac{G^2m_2}{r^3}{\bf{S}}_1\cdot&&
\left[{\bf{v}}_1\times{\bf{n}}\left(2m_1-\frac{1}{2}m_2\right)+{\bf{v}}_2\times{\bf{n}}\left(2m_2\right)\right] \nn\\
+ \frac{Gm_2}{r^2}&&
\left[S_1^{0i}n^i\left(
1-\frac{3}{2}{\bf{v}}_1\cdot{\bf{v}}_2+\frac{3}{2}v_2^2-\frac{3}{2}({\bf{v}}_1\cdot{\bf{n}})({\bf{v}}_2\cdot{\bf{n}})\right) + S_1^{0i}v_2^i\left(-\frac{3}{2}{\bf{v}}_1\cdot{\bf{n}}\right)\right]\nn\\
+\frac{Gm_2}{r}&&\left[\frac{3}{2}\partial_tS_1^{0i}v_2^i\right]-\frac{G^2m_2}{r^3}
S_1^{0i}n^i\left[m_1+2m_2
\right]. 
\eea
This result is to be compared with Eq.~(4.11b) of \cite{Damour:2007nc}.

The NLO SO Lagrangian should be Legendre transformed with respect to the velocities ${\bf{v}}_a$ in order to obtain a NLO SO Hamiltonian that we can compare. We recall that the Legendre transformation with respect to the coordinate conjugate to spin is trivial and may be considered as done once the spin is treated as an independent variable and the Lagrangian is taken with the opposite sign. The part of the Lagrangian required for the obtainment of the NLO SO Hamiltonian $H_{SO}^{NLO}$ reads
\be\label{eq:lreq}  
L = L_N + L_{EIH} - V_{SO}^{LO} + L_{SO}^{NLO},
\ee
where $L_N$ is the Newtonian Lagrangian given by 
\be
L_N = \frac{1}{2}\sum_{a=1}^{2}m_av_a^2 + \frac{Gm_1m_2}{r},
\ee
and $L_{EIH}$ is the Einstein-Infeld-Hoffmann Lagrangian, which is the 1PN correction of the Lagrangian \cite{Goldberger:2004jt}, given by 
\be
L_{EIH}= \frac{1}{8}\sum_{a=1}^{2}m_av_a^4 + \frac{Gm_1m_2}{2r}\left[3v_1^2+3v_2^2-7{\bf{v}}_1\cdot{\bf{v}}_2-({\bf{v}}_1\cdot{\bf{n}})({\bf{v}}_2\cdot{\bf{n}})\right]-\frac{G^2m_1m_2(m_1+m_2)}{2r^2},
\ee
and where $V_{SO}^{LO}$ and $L_{SO}^{NLO}$ are the LO and the NLO spin-orbit couplings, which are given here in Eqs.~(\ref{eq:vlo}) and (\ref{eq:nossc}) respectively. The canonical momenta conjugate to the coordinate velocities ${\bf{v}}_a$ are given by 
\be 
{\bf{p}}_a = \frac{\partial L}{\partial {\bf{v}}_a},
\ee
so that we have
\be 
{\bf{p}}_1 = m_1{\bf{v}}_1 + \frac{m_1v_1^2}{2}{\bf{v}}_1 + \frac{Gm_1m_2}{r}\left(3{\bf{v}}_1-\frac{7}{2}{\bf{v}}_2-\frac{{\bf{v}}_2\cdot{\bf{n}}}{2}{\bf{n}}\right) - 
\frac{G}{r^2}\left(m_2{\bf{S}}_1\times{\bf{n}}+2m_1{\bf{S}}_2\times{\bf{n}}\right) + 
\frac{\partial L_{SO}^{NLO}}{\partial {\bf{v}}_1},
\ee
where we note that here $S^{i0}$ is still considered as an independent degree of freedom, and ${\bf{p}}_2$ is obtained from ${\bf{p}}_1$ by exchanging particle labels $1\leftrightarrow2$ in the above expression. By inverting the relation between conjugate momentum to the coordinate velocity to the desired PN order, we have for the coordinate velocity the following PN expansion in terms of the canonical momenta \cite{Damour:2007nc}
\be\label{eq:vc}
{\bf{v}}_1 = \frac{{\bf{p}}_1}{m_1} - \frac{p_1^2}{2m_1^3}{\bf{p}}_1+\frac{G}{r}\left(-3\frac{m_2}{m_1}{\bf{p}}_1+\frac{7}{2}{\bf{p}}_2+\frac{{\bf{p}}_2\cdot{\bf{n}}}{2}{\bf{n}}\right) + \frac{G}{m_1r^2}\left(m_2{\bf{S}}_1\times{\bf{n}}+2m_1{\bf{S}}_2\times{\bf{n}}\right)- 
\frac{\frac{\partial L_{SO}^{NLO}}{\partial {\bf{v}}_1}}{m_1}. 
\ee
This can be obtained equivalently by using ${\bf{v}}_1 = \frac{\partial H}{\partial {\bf{p}}_1}$ with the relevant Hamiltonian. Again, ${\bf{v}}_2$ is obtained from ${\bf{v}}_1$ by exchanging particle labels $ 1\leftrightarrow2$ in the above expression. Finally, we perform the Legendre transformation, namely, we  substitute in the coordinate velocities using Eq.~(\ref{eq:vc}) in the following expression: 
\be
H = \sum_{a=1,2}{\bf{v}}_a\cdot{\bf{p}}_a - L.
\ee 
Note that the $\frac{\partial L_{SO}^{NLO}}{\partial {\bf{v}}_a}$ terms get canceled and do not contribute here.

The obtained Hamiltonian then takes the following form: 
\be
H = H_O + H_{SO} + H_{SS},
\ee 
where $H_O$ is part of the orbital interaction Hamiltonian, $H_{SO}$ is part of the spin-orbit (linear in spins) Hamiltonian, and $H_{SS}$ is part of the spin-spin (quadratic in spins) sector of the Hamiltonian. Thus, the NLO spin-orbit Hamiltonian we obtain equals 
\bea\label{eq:hso}
H_{SO}^{NLO} = -L_{SO}^{NLO}
&&+  
\frac{Gm_2}{r^2}{\bf{S}}_1\cdot
\left[\frac{{\bf{p}}_1\times{\bf{n}}}{m_1}~\frac{p_1^2}{2m_1^2}-\frac{{\bf{p}}_2\times{\bf{n}}}{m_2}~\frac{p_2^2}{m_2^2} \right] \nn\\ 
&&+  
\frac{G^2m_2}{r^3}{\bf{S}}_1\cdot
\left[\frac{{\bf{p}}_1\times{\bf{n}}}{m_1}\left(7m_1+3m_2\right)+\frac{{\bf{p}}_2\times{\bf{n}}}{m_2}\left(-6m_1-\frac{7}{2}m_2\right) \right] + \left[ 1\leftrightarrow2 \right] \nn\\ 
=\frac{Gm_2}{r^2}{\bf{S}}_1\cdot
&& 
\left[\frac{{\bf{p}}_1\times{\bf{n}}}{m_1}\left(\frac{p_1^2}{2m_1^2}-\frac{{\bf{p}}_1\cdot{\bf{p}}_2}{2m_1m_2}+\frac{p_2^2}{2m_2^2}+\frac{3({\bf{p}}_1\cdot{\bf{n}})({\bf{p}}_2\cdot{\bf{n}})}{2m_1m_2}\right) \right.\nn\\
&&+ 
\left.\frac{{\bf{p}}_2\times{\bf{n}}}{m_2}\left(-\frac{{\bf{p}}_1\cdot{\bf{p}}_2}{m_1m_2}-\frac{3({\bf{p}}_1\cdot{\bf{n}})({\bf{p}}_2\cdot{\bf{n}})}{m_1m_2}\right) +
\frac{{\bf{p}}_1\times{\bf{p}}_2}{m_1m_2}\left(-\frac{{\bf{p}}_1\cdot{\bf{n}}}{2m_1}-\frac{{\bf{p}}_2\cdot{\bf{n}}}{m_2}\right)\right] \nn\\ 
+\frac{G^2m_2}{r^3}{\bf{S}}_1\cdot
&&  
\left[\frac{{\bf{p}}_1\times{\bf{n}}}{m_1}\left(5m_1+\frac{7}{2}m_2\right)+\frac{{\bf{p}}_2\times{\bf{n}}}{m_2}\left(-6m_1-\frac{11}{2}m_2\right) \right] + \nn\\
+ \frac{Gm_2}{r^2}&&
\left[S_1^{i0}n^i\left(
1-\frac{3{\bf{p}}_1\cdot{\bf{p}}_2}{2m_1m_2}+\frac{3p_2^2}{2m_2^2}-\frac{3({\bf{p}}_1\cdot{\bf{n}})({\bf{p}}_2\cdot{\bf{n}})}{2m_1m_2}\right) + S_1^{i0}\frac{p_2^i}{m_2}\left(-\frac{3{\bf{p}}_1\cdot{\bf{n}}}{2m_1}\right)\right]\nn\\
+\frac{Gm_2}{r}&&\left[\frac{3}{2}\partial_tS_1^{i0}\frac{p_2^i}{m_2}\right]-\frac{G^2m_2}{r^3}
S_1^{i0}n^i\left[m_1+2m_2
\right] + \left[ 1\leftrightarrow2 \right].
\eea 
Note the first term appearing in the SSC dependent part of the Hamiltonian, arising from the LO SO interaction. We need to pay special attention and keep track of this $S^{0i}$ term, as it contributes to the NLO SO interaction too.

Now, we are going to eliminate the $S^{0i}$ entries from the Hamiltonian using the covariant SSC in Eq.~(\ref{eq:sscpro}), utilizing Eq.~(\ref{eq:vc}), and the metric for spinning binary black holes in harmonic coordinates, e.g.~as in \cite{Steinhoff:2009hb}. It is well known that this will yield noncanonical variables in the Hamiltonian. However, we can employ noncanonical transformations of variables that relate the variables to canonical ones \cite{Hanson:1974qy}. Therefore, after the elimination with the covariant SSC, we obtain the following Hamiltonian with noncanonical variables:  
\bea\label{eq:hsoredcov}
H_{SO}^{NLO} = 
\frac{Gm_2}{r^2}{\bf{S}}_1\cdot
&& 
\left[\frac{{\bf{p}}_1\times{\bf{n}}}{m_1}\left(\frac{p_1^2}{m_1^2}+\frac{{\bf{p}}_1\cdot{\bf{p}}_2}{m_1m_2}-\frac{p_2^2}{m_2^2}+\frac{3({\bf{p}}_1\cdot{\bf{n}})({\bf{p}}_2\cdot{\bf{n}})}{m_1m_2}\right) \right.\nn\\
&&+ 
\left.\frac{{\bf{p}}_2\times{\bf{n}}}{m_2}\left(-\frac{{\bf{p}}_1\cdot{\bf{p}}_2}{m_1m_2}-\frac{3({\bf{p}}_1\cdot{\bf{n}})({\bf{p}}_2\cdot{\bf{n}})}{m_1m_2}\right) +
\frac{{\bf{p}}_1\times{\bf{p}}_2}{m_1m_2}\left(\frac{{\bf{p}}_1\cdot{\bf{n}}}{m_1}-\frac{{\bf{p}}_2\cdot{\bf{n}}}{m_2}\right)\right] \nn\\ 
+\frac{G^2m_2}{r^3}{\bf{S}}_1\cdot
&&  
\left[\frac{{\bf{p}}_1\times{\bf{n}}}{m_1}\left(6m_1+\frac{13}{2}m_2\right)+\frac{{\bf{p}}_2\times{\bf{n}}}{m_2}\left(-6m_1-\frac{17}{2}m_2\right) \right] + \left[ 1\leftrightarrow2 \right].
\eea
We go on to apply the noncanonical transformations of variables that read
\bea
{\bf{S}}_1 &\to& {\bf{S}}_1 + \frac{p_1^2}{2m_1^2}{\bf{S}}_1 - \frac{{\bf{p}}_1\cdot {\bf{S}}_1}{2m_1^2} {\bf{p}}_1 \equiv {\bf{S}}_1 + \frac{{\bf{p}}_1\times({\bf{S}}_1\times {\bf{p}}_1)}{2m_1^2},\label{eq:scan}\\
{\bf{r}}_1 &\to& {\bf{r}}_1 + \frac{{\bf{S}}_1\times {\bf{p}}_1}{2m_1^2}\left(1-\frac{p_1^2}{4m_1^2}\right)-\frac{Gm_2}{r}\left(\frac{{\bf{S}}_1\times {\bf{p}}_1}{m_1^2}-\frac{3}{2}\frac{{\bf{S}}_1 \times {\bf{p}}_2}{m_1m_2}\right).\label{eq:rcan}
\eea
These transformations are just the generalization of the mapping between covariant and NW SSC variables in flat spacetime \cite{Hanson:1974qy}. The transformation of the spin variable in Eq.~(\ref{eq:scan}) is just similar to that of flat spacetime, whereas for the center of mass coordinate in Eq.~(\ref{eq:rcan}) a higher order PN transformation, than what we had in Eq.~(\ref{eq:covtonw}), generalized for curved spacetime, is required  \cite{Damour:2007nc,Hergt:2010pa}. We argue that now we have arrived at a canonical Hamiltonian.

As was mentioned before, we compare our result with that given by Eq.~(4.11b) of \cite{Damour:2007nc}, or in our notations  
\bea\label{eq:djs}
H_{SO}^{DJS} = \frac{Gm_2}{r^2}{\bf{S}}_1\cdot &&
\left[\frac{{\bf{p}}_1\times{\bf{n}}}{m_1}\left(
\frac{5p_1^2}{8m_1^2}+\frac{3{\bf{p}}_1\cdot{\bf{p}}_2}{4m_1m_2}-\frac{3p_2^2}{4m_2^2}+\frac{3({\bf{p}}_1\cdot{\bf{n}})({\bf{p}}_2\cdot{\bf{n}})}{4m_1m_2}+\frac{3({\bf{p}}_2\cdot{\bf{n}})^2}{2m_2^2}\right) 
\right.\nn\\
&&+ 
\left.\frac{{\bf{p}}_2\times{\bf{n}}}{m_2}\left(
-\frac{{\bf{p}}_1\cdot{\bf{p}}_2}{m_1m_2}-3\frac{({\bf{p}}_1\cdot{\bf{n}})({\bf{p}}_2\cdot{\bf{n}})}{m_1m_2}\right)+\frac{{\bf{p}}_1\times{\bf{p}}_2}{m_1m_2}\left(\frac{3{\bf{p}}_1\cdot{\bf{n}}}{4m_1}-2\frac{{\bf{p}}_2\cdot{\bf{n}}}{m_2}\right)\right]
\nn\\
+\frac{G^2m_2}{r^3}{\bf{S}}_1\cdot &&
\left[\frac{{\bf{p}}_1\times{\bf{n}}}{m_1}\left(\frac{11}{2}m_1+5m_2\right)+\frac{{\bf{p}}_2\times{\bf{n}}}{m_2}\left(-6m_1-\frac{15}{2}m_2\right)\right] + \left[ 1\leftrightarrow2 \right].
\eea 
The difference between the part linear in ${\bf{S}}_1$ in $H_{SO}^{DJS}$ and that of the spin-orbit Hamiltonian obtained here in Eq.~(\ref{eq:hsoredcov}), after the application of the noncanonical transformations in Eqs.~(\ref{eq:scan}) and (\ref{eq:rcan}), i.e.
\be
\Delta H_{SO}^{NLO} = H_{SO}^{DJS} - H_{SO}^{EFT},
\ee
is given by
\bea\label{eq:dso}
\Delta H_{SO}^{NLO} = \frac{Gm_2}{r^2}{\bf{S}}_1\cdot &&
\left[\frac{{\bf{p}}_1\times{\bf{n}}}{m_1}\left(
\frac{{\bf{p}}_1\cdot{\bf{p}}_2}{2m_1m_2}-\frac{p_2^2}{2m_2^2}-\frac{3({\bf{p}}_1\cdot{\bf{n}})({\bf{p}}_2\cdot{\bf{n}})}{2m_1m_2}+\frac{3({\bf{p}}_2\cdot{\bf{n}})^2}{2m_2^2}\right) \right.\nn\\
&&+ 
\left.\frac{{\bf{p}}_1\times{\bf{p}}_2}{m_1m_2}\left(\frac{{\bf{p}}_1\cdot{\bf{n}}}{2m_1}-\frac{{\bf{p}}_2\cdot{\bf{n}}}{m_2}\right)\right]\nn\\
+\frac{G^2m_2}{r^3}{\bf{S}}_1\cdot &&
\left[\frac{{\bf{p}}_2\times{\bf{n}}}{m_2}\left(-\frac{1}{2}m_2\right)\right]. 
\eea 
If the Hamiltonian obtained here is physically equivalent to that of \cite{Damour:2007nc}, i.e.~if it yields the same EOM, there should exist an infinitesimal generator $g$ of a canonical transformation (CT) such that 
\be\label{eq:gH} 
\Delta H_{SO}^{NLO}=\{H,g\}=-\frac{dg}{dt}.
\ee

So let us construct a suitable infinitesimal generator following the discussion of PN canonical transformations in Appendix \ref{app:b} . Such a generator must be a scalar, linear in the spin vector, and constructed out of the spin vector, ${\bf{p}}_1$, ${\bf{p}}_2$, and ${\bf{n}}$. Since the spin already carries $v^1$ in it, we note that there may be three types of generators: $O(G^0p^4)$, $O(G^1p^2)$, and $O(G^2p^0)$. However, only the $O(G^1p^2)$ type contributes appropriate terms. Hence, we consider the following ansatz for the $O(G^1p^2)$ generator: 
\be\label{eq:gen}
g= \frac{Gm_2}{r}{\bf{S}}_1 \cdot \left[g_1\frac{{\bf{p}}_1\times{\bf{p}}_2}{m_1m_2}+\frac{{\bf{p}}_1\times{\bf{n}}}{m_1}\left(g_2\frac{{\bf{p}}_1\cdot{\bf{n}}}{m_1}+g_3\frac{{\bf{p}}_2\cdot{\bf{n}}}{m_2}\right)
+
\frac{{\bf{p}}_2\times{\bf{n}}}{m_2}\left(g_4\frac{{\bf{p}}_1\cdot{\bf{n}}}{m_1}+g_5\frac{{\bf{p}}_2\cdot{\bf{n}}}{m_2}\right)\right].
\ee
This generates both $O(G)$ and $O(G^2)$ terms of the form that appears in the NLO spin-orbit sector:   
$O(G)$ terms from differentiating the coordinates with respect to time, and $O(G^2)$ terms from differentiating the momenta with respect to time, or more precisely $\dot{\bf{p}}$ terms, which become $O(G^2)$ terms upon the use of LO EOM. For example, from the $g_1$ part of the generator in Eq.~(\ref{eq:gen}), we get the following transformations for the canonical variables ${\bf{r}}_1$ and ${\bf{p}}_1$:
\bea
{\bf{r}}_1 &\to& {\bf{r}}_1 - g_1\frac{Gm_2}{r} \frac{{\bf{S}}_1 \times {\bf{p}}_2}{m_1m_2}, \nn \\
{\bf{p}}_1 &\to& {\bf{p}}_1 + g_1\frac{Gm_2}{r^2}~{\bf{S}}_1\cdot\frac{{\bf{p}}_1 \times {\bf{p}}_2}{m_1m_2}~{\bf{n}}.
\eea    

Thus, we plug in Eq.~(\ref{eq:gH}) our ansatz for $g$ from Eq.~(\ref{eq:gen}), and we compare that to Eq.~(\ref{eq:dso}). Comparing $O(G)$ terms gives
\be
g_1=\frac{1}{2},~~g_2=0,~~g_3=-\frac{1}{2},~~g_4=0,~~g_5=0.
\ee
This eliminates all of the $O(G)$ terms, as well as the $O(G^2)$ term from the difference in Eq.~(\ref{eq:dso}), i.e.~finally 
\bea\label{dson}
\Delta H_{SO}^{NLO} = 0
\eea 
in agreement with the canonical result in \cite{Damour:2007nc}.

It is interesting to note that had we used the covariant SSC at the level of the action, rather than considering the temporal spin entries $S^{i0}$ as independent degrees of freedom until their elimination at the Hamiltonian level, we would have arrived at the same Hamiltonian appearing in Eq.~(\ref{eq:hsoredcov}). 
We also note for completeness that the physical equivalence of our result with that of \cite{Damour:2007nc} or \cite{Faye:2006gx} can also be shown by computing and recovering the EOM, using Dirac brackets. However, working with variable transformations is more efficient, which becomes crucial when computing similar or higher order spin corrections.

\section{Conclusions}\label{sec:concl}

In this paper we applied an EFT approach to calculate the NLO gravitational spin-orbit interaction between two spinning compact objects. The NLO spin-orbit interaction was the last NLO conservative spin correction to be computed within the EFT approach, due to its higher complexity with respect to the NLO spin1-spin2 and spin-squared sectors, previously computed for the first time within the EFT approach. The calculation of the NLO SO sector involves the evaluation of 18 Feynman diagrams, 7 of which correspond to one-loop diagrams, including nonstationary cubic self-gravitational interaction, as well as other nonstationary contributions. It is also worth noting that in the NLO SO sector one cannot avoid having acceleration dependent terms, and using EOM to eliminate them. The NRG field decomposition, first applied in the computation of the NLO spin1-spin2 interaction, in terms of which the EFT calculation is carried out here, facilitates the calculation considerably. We recall that the NRG field decomposition in the harmonic gauge has many advantages, such as simple propagators for the scalar and vector fields, which dominate in the SO interaction, and the vanishing of mixed 2-point functions. In particular, the NRG field decomposition makes the treatment of the cubic self-gravitational interaction relatively simple, so that there is no need to rely on automated computations at all. In addition, the NRG field decomposition illustrates the coupling hierarchy of the different gravitational field components to the spin and mass sources, which is helpful in the construction of Feynman diagrams. 

However, the main obstacle in the calculation of the PN spin corrections is the treatment of SSC. Therefore, the fact that, unlike the LO spin1-spin2 or spin-squared effects the LO spin-orbit effect already involves the use of SSC presents a major complication in this sector. We showed here that already at the LO SO sector, this results in ambiguities related with different choices of SSC, which can be resolved in terms of noncanonical changes of variables. Moreover, the NLO SO sector requires the application of SSC at higher orders, whereas SSCs at higher orders were not considered until recently, and they should be further explored. Finally, we have shown here explicitly how to relate the EFT derived spin results to the canonical results obtained with the ADM Hamiltonian formalism. This is done, similarly to the LO SO sector, by using noncanonical transformations, as well as canonical transformations at the level of the Hamiltonian. We note, however, that it is more difficult to apply on the Hamiltonian noncanonical transformations compared to canonical ones. Yet, working with transformations of variables is more efficient than resorting to the EOM and Dirac brackets, and that becomes crucial when unknown higher order spin corrections are to be approached in the future.
 
\section*{ACKNOWLEDGMENTS}

I am grateful to Barak Kol for his continuous support and encouragement. I would like to thank Michael Smolkin and Shmuel Elitzur for pleasant discussions. I also thank Jan Steinhoff for constructive correspondence. Special thanks to Adam Schwimmer for hospitality at Weizmann Institute. This research is supported by the Israel Science Foundation Grant No.~607/05, and by the German Israeli Project Cooperation Grant DIP H.52.

\appendix
\section{Dimensional regularization and loop integrals}\label{app:a}

Throughout the computation of the contributing Feynman diagrams, we encounter two types of momentum integrals that need to be evaluated: Fourier integrals that arise from the Fourier transforms of the propagators, and one-loop integrals, which arise from the cubic gravitational interaction. Both types of integrals are evaluated using dimensional regularization \cite{Collins:1984xc}. In order to evaluate the Fourier integrals, one should use the d-dimensional master formula given by 
\be\label{eq:ft}
\int \frac{d^d\bf{k}}{(2\pi)^d}\frac{e^{i\bf{k}\cdot\bf{r}}}{({\bf{k}}^2)^\alpha}=\frac{1}{(4\pi)^{d/2}}\frac{\Gamma(d/2-\alpha)}{\Gamma(\alpha)}\left(\frac{{\bf{r}}^2}{4}\right)^{\alpha-d/2}.
\ee
This formula can easily be derived using Schwinger parameters \cite{Collins:1984xc}. From this master formula, we obtain the following required Fourier integrals:
\bea
\int \frac{d^3\bf{k}}{(2\pi)^3}\frac{e^{i\bf{k}\cdot\bf{r}}}{k^2}=&&\frac{1}{4\pi|\bf{r}|},\label{eq:ft1}\\
\int \frac{d^3\bf{k}}{(2\pi)^3}\frac{e^{i\bf{k}\cdot\bf{r}}}{k^4}=&&-\frac{|\bf{r}|}{8\pi},\label{eq:ft2}\\
\int \frac{d^3\bf{k}}{(2\pi)^3}\frac{e^{i\bf{k}\cdot\bf{r}}}{k}=&&\frac{1}{2\pi^2|{\bf{r}}|^2}\label{eq:ft0.5}\\
\int \frac{d^3\bf{k}}{(2\pi)^3}\frac{e^{i\bf{k}\cdot\bf{r}}k^ik^j}{k^3}=&&\frac{1}{2\pi^2|{\bf{r}}|^2}\left(\delta_{ij}-2n_in_j\right).\label{eq:ft1.5}
\eea

The loop integrals required for the computation of the Feynman diagrams that contain three-graviton vertices are given by
\bea
\int \frac{d^3\bf{k}}{(2\pi)^3}\frac{k^i}{{\bf{k}}^2({\bf{k}+\bf{q}})^2}= && -\frac{q^i}{16q},\label{eq:1l1}\\
\int \frac{d^3\bf{k}}{(2\pi)^3}\frac{k^ik^j}{{\bf{k}}^2({\bf{k}+\bf{q}})^2} = &&\frac{1}{64}\left(\frac{3q^iq^j}{q}-\delta^{ij}q\right).\label{eq:1l2}
\eea
These integrals can be obtained using both Feynman and Schwinger parameters \cite{Collins:1984xc}.

\section{PN canonical transformations}\label{app:b} 

Let us elaborate on the statement made by Eq.~(\ref{eq:gH}) and derive it here. Recall that our Lagrangian is acceleration dependent, due to its high PN order (2PN order and beyond). We know that for a Lagrangian of the form $L=L(q,\dot{q},t)$, i.e.~a standard Lagrangian, which depends on some coordinate $q$ and velocity $\dot{q}$, a total time derivative of an arbitrary function $F(q,t)$, which depends on the coordinate $q$ and time only, can be added to the Lagrangian $L$, such that $L(q,\dot{q},t)+\frac{d}{dt}F(q,t)$ is physically equivalent to $L$, i.e.~it yields the same EOM as $L$. Similarly, it is easy to show that for an acceleration dependent Lagrangian of the form $L=L(q,\dot{q},\ddot{q},t)$, a total time derivative of an arbitrary function  $F(q,\dot{q},t)$, which depends also on the velocity $\dot{q}$, can be added to the Lagrangian $L$, such that $L(q,\dot{q},\ddot{q},t)+\frac{d}{dt}F(q,\dot{q},t)$ yields the same EOM as $L$.

An addition of such a total time derivative 
to the Lagrangian corresponds to applying the infinitesimal CTs given by
\bea \label{eq:ict}
q &\to& q + d\alpha\frac{\partial}{\partial p}g(q,p),\nn\\ 
p &\to& p - d\alpha\frac{\partial}{\partial q}g(q,p),
\eea
on the canonical variables $(q,p)$ of the Hamiltonian, where $d\alpha$ is an infinitesimal of the transformation parameter, and $g = g(q,p)$ is the infinitesimal generator of the transformations. If one would like to make an addition to the $\frac{n}{2}$PN part of the Hamiltonian, $g$ should be of the same PN order, or more precisely it should scale as $v^n t$. The generator $g$ should be given in terms of our canonical variables, and it should not depend on time explicitly, since our PN Hamiltonian does not. It can be directly checked that the mappings in Eq.~(\ref{eq:ict}) describe a canonical transformation from the so-called ``direct conditions'' for restricted canonical transformations \cite{Goldstein:1980}, i.e.~CTs that do not contain time explicitly. Note that since $g\sim v^n$, the canonical momenta $p$ that will appear in $g$ will effectively be the Newtonian momenta, i.e.~$p=mv$. 

Now, let $A$ be some function of the canonical variables, which does not depend on time explicitly. Then, it follows that $dA=d\alpha\{A,g\}$ \cite{Goldstein:1980}, so that the change in $A$ due to a finite CT is given by 
\be
\Delta A = \alpha \{A,g\} + \frac{\alpha^2}{2!}\{\{A,g\},g\} + \frac{\alpha^3}{3!}\{\{\{A,g\},g\},g\} + \cdots.
\ee
Since $g$ scales as $v^n$, all orders beyond linear in $\alpha$ do not contribute to the considered PN order, so to that order we have $\Delta A = \alpha \{A,g\}$. From this we find that the change in the Hamiltonian due to the finite CTs, which are of the form 
\bea \label{eq:ct}
q &\to& q + \alpha\frac{\partial g}{\partial p},\nn\\
p &\to& p - \alpha\frac{\partial g}{\partial q},
\eea
is given by
\be
\Delta H = \alpha \{H,g\} = - \alpha \frac{dg}{dt},
\ee
since our $H$ and $g$ does not depend on time explicitly. Thus, we have arrived at Eq.~(\ref{eq:gH}), up to a numerical coefficient, which is the finite transformation parameter.

\end{document}